\documentclass[11pt]{article}

\usepackage{amsmath, amsthm}
\usepackage{amsmath, amsfonts}
\usepackage{amsmath, amssymb}
\usepackage{amsmath}
\usepackage{graphics}
\usepackage{graphicx}
\usepackage{color}
\usepackage{hyperref}
\usepackage[all]{xy}

\newtheorem{theorem}{Theorem}[subsection]

\newtheorem{proto-definition}[theorem]{Proto-Definition}
\newtheorem{pseudo-definition}[theorem]{Pseudo-Definition}
\newtheorem{definition-lemma}[theorem]{Definition/Lemma}
\newtheorem{definition-explanation}[theorem]{Definition/Explanation}
\newtheorem{explanation-definition}[theorem]{Explanation/Definition}
\newtheorem{definition-fact}[theorem]{Definition/Fact}
\newtheorem{definition-notation}[theorem]{Definition/Notation}
\newtheorem{definition-conjecture}[theorem]{Definition/Conjecture}
\newtheorem{definition-theorem}[theorem]{Definition/Theorem}
\newtheorem{lemma}[theorem]{Lemma}
\newtheorem{lemma-definition}[theorem]{Lemma/Definition}

\newtheorem{remark}[theorem]{\it Remark}
\newtheorem{remark-notation}[theorem]{\it Remark/Notation}

\newtheorem{application-lemma}[theorem]{Application/Lemma}

\newtheorem{example-definition}[theorem]{Example/Definition}

\newtheorem{definition-prototype}[theorem]{Definition-Prototype}

\numberwithin{equation}{subsection}

\newtheorem{stheorem}{Theorem}[section]
\newtheorem{sdefinition}[stheorem]{Definition}
\newtheorem{sproto-definition}[stheorem]{Proto-Definition}
\newtheorem{spseudo-definition}[stheorem]{Pseudo-Definition}
\newtheorem{sdefinition-lemma}[stheorem]{Definition/Lemma}
\newtheorem{sdefinition-explanation}[stheorem]{Definition/Explanation}
\newtheorem{sexplanation-definition}[stheorem]{Explanation/Definition}
\newtheorem{sdefinition-fact}[stheorem]{Definition/Fact}
\newtheorem{sdefinition-notation}[stheorem]{Definition/Notation}
\newtheorem{sdefinition-conjecture}[stheorem]{Definition/Conjecture}
\newtheorem{sdefinition-theorem}[stheorem]{Definition/Theorem}

\newtheorem{slemma-definition}[stheorem]{Lemma/Definition}

\newtheorem{scorollary}[stheorem]{Corollary}

\newtheorem{sremark-notation}[stheorem]{\it Remark/Notation}

\newtheorem{sapplication-lemma}[stheorem]{Application/Lemma}

\newtheorem{sexample}[stheorem]{Example}
\newtheorem{sexample-definition}[stheorem]{Example/Definition}

\newtheorem{sdefinition-prototype}[stheorem]{Definition-Prototype}

\newtheorem{ssproto-definition}[sstheorem]{Proto-Definition}
\newtheorem{sspseudo-definition}[sstheorem]{Pseudo-Definition}
\newtheorem{ssdefinition-lemma}[sstheorem]{Definition/Lemma}
\newtheorem{ssdefinition-explanation}[sstheorem]{Definition/Explanation}
\newtheorem{ssexplanation-definition}[sstheorem]{Explanation/Definition}
\newtheorem{ssdefinition-fact}[sstheorem]{Definition/Fact}
\newtheorem{ssdefinition-notation}[sstheorem]{Definition/Notation}
\newtheorem{ssdefinition-conjecture}[sstheorem]{Definition/Conjecture}
\newtheorem{ssdefinition-theorem}[sstheorem]{Definition/Theorem}

\newtheorem{sslemma-definition}[sstheorem]{Lemma/Definition}

\newtheorem{ssremark-notation}[sstheorem]{\it Remark/Notation}

\newtheorem{ssapplication-lemma}[sstheorem]{Application/Lemma}

\newtheorem{ssexample-definition}[sstheorem]{Example/Definition}

\newtheorem{ssdefinition-prototype}[sstheorem]{Definition-Prototype}

 \newcommand{\Exp}{\mbox{\rm Exp}\,}

\newcommand{\Id}{\mbox{\it Id}\,}

\newcommand{\Image}{\mbox{\it Im}\,}

\newcommand{\dil}{\mbox{\rm dil}\,}
\newcommand{\dimless}{\,\mbox{\scriptsize\rm dimless}}

\newcommand{\eapprxn}{\mbox{\scriptsize\rm $\varepsilon$-apprxn}\,}

\newcommand{\phys}{\mbox{\scriptsize\rm phys}}

\newcommand{\word}{\mbox{\scriptsize\rm word}}

\begin{document}

\enlargethispage{24cm}

\begin{titlepage}

$ $

\vspace{-1.5cm} 

\noindent\hspace{-1cm}
\parbox[t]{10cm}{\small November 1995 \\
                   Revised: September 2015}\
   \hspace{4cm}\
   \parbox[t]{6cm}{1509.03895 [gr-qc]}

\vspace{2cm}

\centerline{\large\bf
 Quantum fluctuations, conformal deformations, and Gromov's topology}
\vspace{1ex}
\centerline{\large\bf
 --- Wheeler, DeWitt, and Wilson meeting Gromov}

\vspace{3em}

\centerline{\large
  Chien-Hao Liu}

\vspace{4em}

\begin{quotation}
\centerline{\bf Abstract}

\vspace{0.3cm}

\baselineskip 12pt  
{\small
The moduli space of isometry classes of Riemannian structures on a smooth manifold was emphasized 
 by J.A.\ Wheeler in his superspace formalism of quantum gravity. 
A natural question concerning it is:
 {\it What is a natural topology on such moduli space that reflects best quantum fluctuations of the geometries
  within the Planck's scale?}
This very question has been addressed by B.\ DeWitt and others.
In this article we introduce Gromov's {\it $\varepsilon$-approximation topology}
 on the above moduli space for a closed smooth manifold.
After giving readers some feel of this topology,
 we prove that  each conformal class in the moduli space is dense with respect to this topology.
Implication of this phenomenon to quantum gravity is yet to be explored.
When going further to general metric spaces,
 Gromov's {\it geometries-at-large-scale} based on his topologies
  remind one of K.\ Wilson's {\it theory of renormalization group}.
We discuss some features of both and
 pose a question on whether both can be merged into a single unified theory.
} 
\end{quotation}

\vspace{12em}

\baselineskip 12pt
{\footnotesize
\noindent
{\bf Key words:} \parbox[t]{14cm}{superspace,
 $\varepsilon$-approximation topology, conformal deformation, theory space, renormalization group
 }} 

 \bigskip

\noindent {\small MSC number 2010: 83E05, 53C20; 83C45, 54E35, 81T17
} 

\bigskip

\baselineskip 10pt
{\scriptsize
\noindent{\bf Acknowledgements.}
 I would like to thank
   Bill Thurston
     for introducing Gromov's work to me and his guidance in several enlightening discussions;
   Neil Turok
     whose course deepens my interest in gravity;
   Orlando Alvarez
     for his generous advices in physics and
	      teaching me Wilson's work on Renormalization Group among other things;
   Martin Bridson
     who went through the detail of the Main Theorem and commented on it;
 Thomas Curtright,  Jerrold Marsden, John Mather, Peter Sarnak
    for discussions and remarks;
 Hung-Wen Chang, Ling-Miao Chou, Lang-Fang Wu
    for discussions and encouragements.
 Special thanks to
   Inkang Kim, Ayelet Lindenstrauss, Robert McCann, Arlie Petter, James Tanton, Ann Willman, Wenping Zhang
      for being present in my defense;
 [(added 2015) and
   Ling-Miao Chou for numerous comments that improve the illustrations$^{\ast}$ 
 The mastery flute performance of  Emmanuel Pahud on J.S.\ Bach
   accompanies the retyping of the original manuscript and the redrawing of all the figures.]
   
 \noindent ------------------------------------

 \noindent
 $^{\ast}$
 This version contains simplified illustrations to limit the file size
 for the submission to arXiv.
 For complete illustrations, readers are referred to the version,
 posted one day behind,  in
  $\;$\verb=https://www.researchgate.net/profile/Chien-Hao_Liu/=$\;$                                            
 } 

\end{titlepage}

\newpage

\begin{titlepage}

$ $

\vspace{12em}

\bigskip

\centerline{\small\it
 Dedicated to Professor Neil Turok}
\centerline{\small\it
 for his generous time toward my questions on gravity and cosmology}
\centerline{\small\it
 while I was still hesitating about which path to take at Princeton.}

%
%

\end{titlepage}


\newpage
$ $

\vspace{-3em}

\centerline{\sc
 Conformal Deformations of Riemannian Metrics
 } %

\vspace{2em}


\begin{flushleft}
{\Large\bf 0. Introduction and outline}
\end{flushleft}
The moduli space of isometry classes of Riemannian structures on a smooth manifold was emphasized
 by J.A.\ Wheeler in his superspace formalism of quantum gravity.
A natural question concerning it is:
 {\it What is a natural topology on such moduli space that reflects best quantum fluctuations of the geometries
  within the Planck's scale?}
This very question has been addressed by B.\ DeWitt and others.

In this article we introduce Gromov's {\it $\varepsilon$-approximation topology}
 on the above moduli space for a closed smooth manifold.
After giving readers some feel of this topology,
 we prove the main theorem (Theorem 2.1) in this article which states that
   \begin{itemize}
    \item[]\it
	 Given
	    two metric tensors $ds_0^2$, $ds_1^2$ on a smooth closed $n$-manifold $M$, with $n$ arbitrary, and
		any scale $\varepsilon_0>0$,
     there exists a positive smooth function $\Omega$ on $M$ such that
	    $(M,ds_0^2)$ and $(M,\Omega^2ds_1^2)$ are $\varepsilon_0$-close to each other.	
   \end{itemize}
An immediate consequence (Corollary 2.1) of this is that
 {\it each conformal class in this moduli space is dense
         with respect to Gromov's $\varepsilon$-approximation topology}.
Implication of this phenomenon to quantum gravity is yet to be explored.

When going further to general metric spaces,
 Gromov's {\it geometries-at-large-scale} based on his topologies
  remind one of K.\ Wilson's {\it theory of renormalization group}.
We discuss some features of both and
 pose a question on whether both can be merged into a single unified theory.

\bigskip

\begin{flushleft}
{\bf Outline}
\end{flushleft}
{\small
 \baselineskip 12pt  
 \begin{itemize}       	
   \item[1]
	 Gromov's topology on Wheeler's superspace
	  \vspace{-.6ex}
	  \begin{itemize}
	   \item[\Large$\cdot\;$]
         Wheeler's superspace and DeWitt's question		

	   \item[\Large$\cdot\;$]	
         Gromov's $\varepsilon$-approximation topology
						
       \item[\Large$\cdot\;$]
         Compatibility with fluctuations from quantum gravity
		
	   \item[\Large$\cdot\;$]
         An example -- the case of closed $2$-manifolds
		
       \item[\Large$\cdot\;$]
         A further question
	  \end{itemize}
   	
   \item[2]
	 Main Theorem on conformal deformations
	  \vspace{-.6ex}
	  \begin{itemize}
	   \item[2.1]
         Proof of the theorem

	   \item[2.2]	
         Proof of Lemma 2.1.1
						
       \item[2.3]
	     Proof of Lemma 2.1.2
		 \begin{itemize}
		   \item[2.3.1]
		    Basic bounds and Euclidean approximations

		   \item[2.3.2]	
            A Lipschitz anti-tilt homeomorphism on $\Delta_{(l,R)}$

		   \item[2.3.3]	
            A Lipschitz radial-squeeze homeomorphism in $\Delta_{(l,R)}$

		   \item[2.3.4]	
            Conjugation back to $K$

		   \item[2.3.5]	
            Application to $\Gamma$
         \end{itemize}
	  \end{itemize}
   	
	\item[3]
     Another question -- Toward a ``Gromov-Wilson Theory"?
	  \vspace{-.6ex}
	  \begin{itemize}
	    \item[\Large$\cdot\;$]
         Wilson's renormalization group theory

	    \item[\Large$\cdot\;$]	
         Gromov's geometry-at-large-scale

        \item[\Large$\cdot\;$]
         Wilson $+$ Gromov $=?$
	  \end{itemize}
 \end{itemize}
} 

\newpage

\section{Gromov's topology on Wheeler's superspace}

In this section
 we explain why Gromov's $\varepsilon$-approximation topology on the moduli space of metric spaces
 may be a good/meaningful (resp.\  starting) choice for the topology on a superspace
  in the study of general relativity and gravity (resp.\ before further refinement).

\bigskip

\begin{flushleft}
{\bf Wheeler's superspace and DeWitt's question}
\end{flushleft}		
In his superspace\footnote{Unfamiliar readers
          should be aware that there have been two completely different meanings assigned to the word
                ``superspace" in physics.
          In historical order, it means :
		  (1)
	           {\it Superspace} {\sl in the context of general relativity and gravity} :
	            The moduli space of metric spaces of a certain kind.
	            In other words, it is a ``space of all spaces".	
           (2)				
	           {\it Superspace} {\sl in the context of quantum field theory and particle physics} :
	            A space whose coordinates (or more precisely function-ring) are ${\Bbb Z}/2$-graded.
	            In other others, it is a space that contains
	                  both the bosonic degrees of freedom and fermionic degrees of freedom.
        Definition (1) is taken throughput this note.
      }
  formalism of general relativity,
 John Wheeler interprets Einstein's equations as a dynamical law for $3$-dimensional Riemannian manifolds
  ([Wh1], [Wj2], [Wh3], [Wh4], [Wh5]).
Under  such setting,
  one of the fundamental objects that one would look at
    is the arena where the dynamics takes place,
	i.e.\ the moduli space of isometry classes of $3$-dimensional Riemannian manifolds.
 An unavoidable question, which was posed by 	Bryce DeWitt ([DeW1], [DeW2], [DeW3]) and others, is then:
   \begin{itemize}
    \item[{\bf Q.}]\it
     What is a natural topology on this moduli space
	  that best reflects quantum fluctuations of the geometry within the Planck's scale?	
   \end{itemize}
 Cf.\ {\sc Figure}~1-1.
     %

 \begin{figure} [htbp]
  \bigskip
  \centering

  \includegraphics[width=0.8\textwidth]{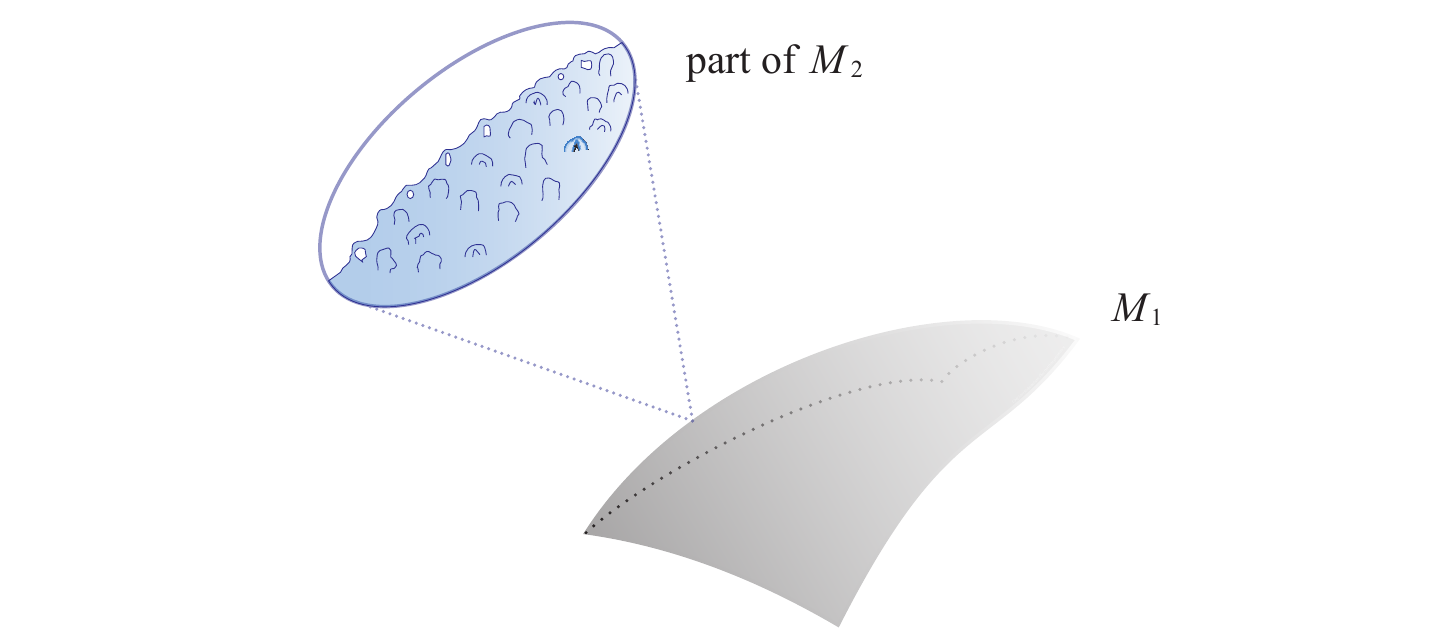}

  \bigskip
  \bigskip
  \centerline{\parbox{13cm}{\small\baselineskip 12pt
   {\sc Figure}~1-1.
    When zooming into the Planck's scale, the topology and geometry of a space fluctuate constantly.
	{\bf Q.} {\it As abstract metric  spaces,
	     should the large-scale geometry $M_1$ and the Planck-scale geometry $M_2$ be regarded
		 as being close to each other in Wheeler's superspace?}
     (Figure adapted from [Wh4: Sek.\ 18: Abbild 7]. )		 	 	
	In the figure, a tiny portion of $M_2$ is revealed through zooming in.
       }}
  \bigskip
 \end{figure}


\begin{flushleft}
{\bf Gromov's $\varepsilon$-approximation topology}
\end{flushleft}
In his pioneering work on geometric group theory,
 Mikhail Gromov introduces several notions of {\it quasi-isometries} between metric spaces
 ([C-E-G], [Gr1], [Gr2, [Gr3], [Gr4]]).
Among them is the following notion of {\it $\varepsilon$-approximation} :

\bigskip

\begin{sdefinition} {\bf [$\varepsilon$-approximation].}
 Let
   $(X,d)$ and $(Y,d^{\prime})$ be two metric spaces and
   $\varepsilon$ be a non-negative real number.
 An {\it $\varepsilon$-approximation} from $(X,d)$ to $(Y,d^{\prime})$
  is a multi-valued surjective map\footnote{There
                      are several equivalent versions of the notion of $\varepsilon$-approximation in the literature.
					 Here we select one that treats it as a multi-valued map.
					 This gives us a sense of direction in such an approximation and is more convenient to use later.}
    $$
	  R_{\varepsilon}\;:\;
	   (X,d)\; \longrightarrow\; (Y,d^{\prime})	
	$$
  such that
    $$
      |d(x_1,x_2)-d^{\prime}(y_1,y_2)|\; \le \; \varepsilon
    $$	
 for all $x_i\in X$, $y_i\in R_{\varepsilon}(x_i)$, $i=1,\,2$.	
\end{sdefinition}

\bigskip

Two metric spaces will be said to be {\it $\varepsilon$-close}
 if there exists an ${\varepsilon}$-approximation between them.
This then defines a pseudo-distance $d_{\eapprxn}$ between metric spaces $X$ and $Y$ by taking
 $$
   d_{\eapprxn}(X,Y)\; :=\;
     \inf\{\varepsilon\,|\, \mbox{there exists an $\varepsilon$-approximation between $X$ and $Y$}\}\,.
 $$
This pseudo-distance defines in turn a (generally non-Hausdorff) topology
   on the space of isometry classes of metric spaces in the category in question,
  which we shall call the {\it Gromov's ($\varepsilon$-approximation) topology}.
As long as the purpose of this work is concerned,
 it is both crucial and natural to ask :
  \begin{itemize}
   \item[{\bf Q.}]\it
    Does this topology on a superspace manifest
  	  the nature of the topological and geometrical fluctuations of spaces from the aspect of quantum gravity?
  \end{itemize}
  
Cf.\ {\sc Figure}~1-2.
     %

 \begin{figure} [htbp]
  \bigskip
  \centering

  \includegraphics[width=0.8\textwidth]{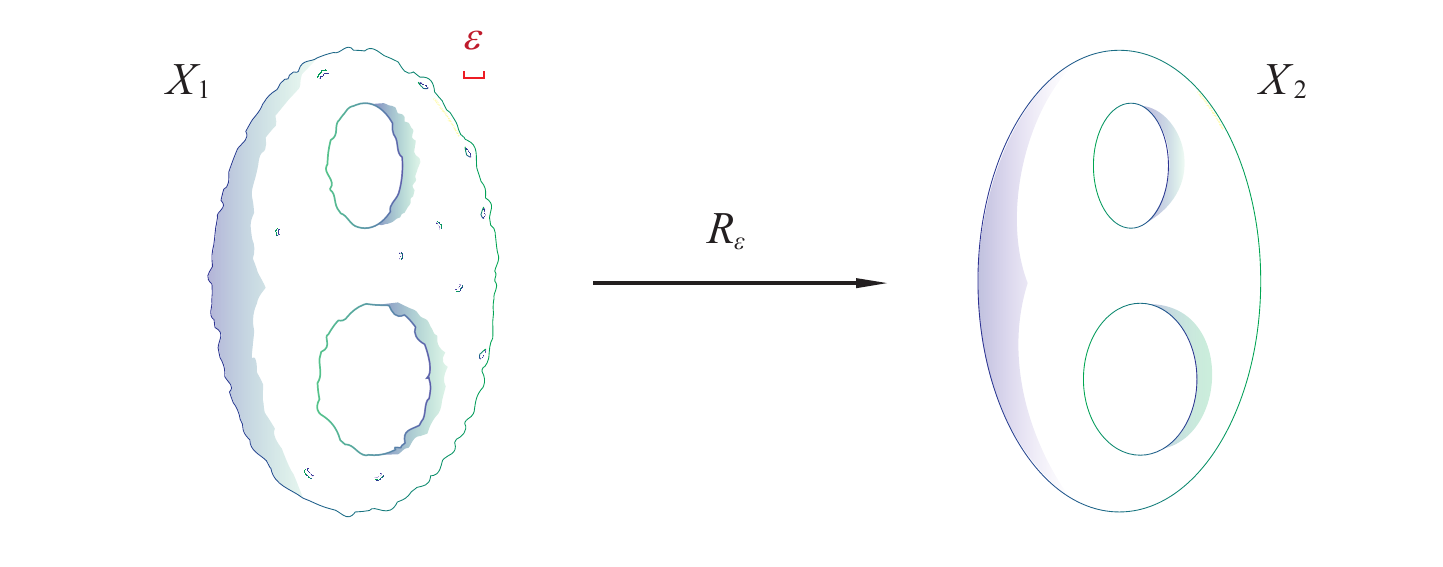}

  \bigskip
  \bigskip
  \centerline{\parbox{13cm}{\small\baselineskip 12pt
   {\sc Figure}~1-2.
    Two metric spaces $X_1$ and $X_2$ that are related by an $\varepsilon$-approximation
	 $R_{\varepsilon}:X_1\rightarrow X_2$
	 could be drastically different in both topology and geometry
 	at the scale smaller than $\varepsilon$.
	However, such differences become vague when one looks at them at a scale that is large
	 with respect to $\varepsilon$.
		 }}
  \bigskip
 \end{figure}	

						
\begin{flushleft}
{\bf Compatibility with fluctuations from quantum gravity}
\end{flushleft}		
In Wheeler's picture, the topological and the geometrical fluctuations of a space happen habitually 
 when one probes the geometry at the Planck's scale.
However,
 since classical physics (including gravitation) is at  a scale much larger than the scale for such fluctuations
      and is very stable,
 it must be quite insensitive to such fluctuations.
Consequently,  one would expect that
 the new geometries (supported in general by new topologies) arising from such quantum fluctuations
  are indistinguishable from the original one
 if one ``looks at or measure them" with a scale/unmarked ruler much larger than the scale of fluctuations.
The profundity of Gromov's definition is that :
 \begin{itemize}
  \item[\LARGE $\cdot$]\it
   It makes precise and does capture the above feature one wishes for from quantum gravity.
 \end{itemize}
  
When compared with some other known topologies on the moduli space of geometries [Fi],
 Gromov's notion distinguishes itself by two features :
 \begin{itemize}
  \item[(1)]\it
   It sets a scale in the definition for closeness/nearbyness in the moduli space;

  \item[(2)]\it
   It treats topological fluctuations as fairly as geometrical fluctuations.
 \end{itemize}
For puritans, one may even say that :
 \begin{itemize}
  \item[\LARGE $\cdot$]
   Gromov's topology {\it defines} what it means by
    ``{\it topological and geometrical fluctuations within scale $\varepsilon$}"
	precisely and mathematically as
	{\it a random choice of the geometry (with possibly a different topology from the original one)
	in the $\varepsilon$-neighborhood of the original geometry
	in the moduli space (i.e.\ superspace) in question}.
 \end{itemize}

\bigskip

\begin{flushleft}
{\bf An example -- the case of closed $2$-manifolds}
\end{flushleft}
To have a concrete feel of Gromov's topology,
 let us take a brief glance at what the moduli space should look like
  for the case of closed Riemannian $2$-manifolds.
Recall that the topology of closed $2$-manifolds $\Sigma$ are classified by their orientability
 and Euler characteristic $\chi(\Sigma)$.
They can all be obtained by connected-summing a $2$-sphere $S^2$
   with finitely many $2$-torus $T^2$ (cf.\ handles)
       and at most two ${\Bbb R}P^2$ (cf.\ crosscaps).
Denote
   the latter pair of numbers by $(h,c)$ and
   let ${\cal M}_{(h,c)}$ be the moduli space of isometry classes of Riemannian manifolds
      supported by the corresponding topology and
   let ${\cal M}_2$ be the union of all ${\cal M}_{(h,c)}$,
   which forms the whole moduli space in the present case.
   
One can show that, with Gromov's topology,
 ${\cal M}_2$ is stratified by ${\cal M}_{(h,c)}$'s
  in a way that follows the order of complexity of the topology of $\Sigma$'s indicated by $\chi(\Sigma)$,
  as illustrated below :
 $$
   \xymatrix{
    &   \vdots    & \hspace{.8em}\vdots\rule[-1ex]{0ex}{1em}\hspace{.8em}
	          & \vdots\rule[-1ex]{0ex}{1em}  \\
	&{\cal M}_{(h,2)} \ar[ru]
	          & {\cal M}_{(h,1)}\ar[l]
	          & {\cal M}_{(h,0)}\ar[l] \ar[u]\\
	&   \hspace{.8em}\vdots\rule[-1ex]{0ex}{1em}\hspace{.8em} \ar[ru]
	           & \hspace{.8em}\vdots\rule[-1ex]{0ex}{1em}\hspace{.8em} \ar[l]
	           & \hspace{.8em}\vdots\rule[-1ex]{0ex}{1em}\hspace{.8em} \ar[l] \ar[u] \\     		 
	&  {\cal M}_{(1,2)} \ar[ur]
	           & {\cal M}_{(1,1)}\ar[l]
	           & {\cal M}_{(1,0)}\ar[u] \ar[l] \\
	\,&  {\cal M}_{(0,2)} \ar[ur]
		       & {\cal M}_{(0,1)} \ar[l]
		       & {\cal M}_{(0,0)} \ar[l] \ar[u]  &,
      }
 $$
where the arrows `$\longrightarrow$' in the diagram means `in the boundary of'.
The strata relations that are generated by the transitivity of the relation $\rightarrow$ are omitted in the diagram.
In particular, any $\varepsilon$-neighborhood of a $\Sigma_0\in {\cal M}_2$
  contains some other $\Sigma$'s of different (more complicated) topology.
  
The above result follows from two observations, which we shall only state without proof :
 (a point in ${\cal M}_2$ is alled a {\it $2$-geometry})
 \begin{itemize}
  \item[(1)]
   [{\it collapsing of fundamental group}]\\
   Let $\Sigma_i$, $i\in {\Bbb Z}_{\ge 1}$
        be a convergent sequence of  $2$-geometries in ${\cal M}_2$
    	that lie completely in the stratum associated to a $2$-manifold $\Sigma$.
   Let $\Sigma_{\infty}$ be the limit.
   Then, the findamental group $\pi_1(\Sigma_{\infty})$ is a quotient of $\pi_1(\Sigma)$.
   
  \item[(2)]
   [{\it non-orientable unapproximable by fixed orientable}]\\
   Let
     $\Sigma_{\infty}\in{\cal M}_2$  be non-orientable and
	 $\Sigma_i$, $i\in {\Bbb Z}_{\ge 1}$
    	 be a sequence of orientable of $2$-geometries in ${\cal M}_2$.
   Then, $\lim_{i\rightarrow \infty}\chi(\Sigma_i)=-\infty$.	
 \end{itemize}
 
 One can also show that
  \begin{itemize}
   \item[\LARGE $\cdot$]
    Any path-metric space supported by a closed $2$-manifold $\Sigma$
	 can be approximated arbitrarily closely by some (generally very bumpy) Riemannian metric space
	 supported on the same $2$-manifold $\Sigma$.
  \end{itemize}
 This gives another illustration that a geometry can be disguied when observed only up a certain scale.

\bigskip

\begin{flushleft}		
{\bf A further question}
\end{flushleft}
Previous discussions and lessons from the $2$-dimensional case suggest that
 \begin{itemize}
  \item[\LARGE $\cdot$]
   A natural arena for quantum gravity is the moduli space of isometry classes of certain type of 
    path-metric spaces, in which the submoduli space of Riemannian manifolds of that type is
    dense-at-the-Planck-scale with respect to the Gromov's topology.
 \end{itemize}
Both the path-metric analogue of Riemannian geometry and the generalization of general relativity
 to such spaces have been investigated in the literature (e.g.\ [A-B-N], [A-M-W], [Br], [Gr3], [Mi], [Re]).
One would next want to
  construct a compatible action defined as a functional on the path space of this moduli space
  in order to have the dynamics for the geometry.
Ideally one would like to construct such an action
 so that its extremals are close to the extremals of the Hilbert-Einstein action
  on the path space of the submoduli space of Riemannian manifolds involved.
In view of the fact that
 neither curvatures nor the volume is a continuous function on the moduli space with the Gromov's topology,
it seems extremely difficult to realize this goal.

\bigskip

With these background, let us now turn to the core --- the Main Theorem --- of this note in the next section.

\bigskip
   	
\section{Main Theorem on conformal deformations}

Conformal deformations and conformal geometries have been of interest to both mathematicians and physicists.
The following theorem on conformal deformations,
  though perhaps surprising at the first sight, indicates its ability to distort/warp shapes of Riemannian manifolds.

\bigskip
\bigskip

\begin{stheorem} {\bf [conformal deformation].}
 Let
  \begin{itemize}
    \item[\LARGE $\cdot$]
     $M$ be a closed smooth $n$-manifold,
	
	\item[\LARGE $\cdot$]
     $ds_0^2$ and $ds_1^2$ be two Riemannian metrics on $M$,
	
	\item[\LARGE $\cdot$]
     $\varepsilon_0$ be any given positive real number.
  \end{itemize}
 Then there exists a positive smooth function $\Omega$ on $M$ such that
    $(M,ds_0^2)$ and $(M, \Omega^2ds_1^2)$ are $\varepsilon_0$-close to each other.
\end{stheorem}

\bigskip

As an immediate consequence, one has the following density property,
 whose implication to quantum gravity is yet to be explored :

\bigskip

\begin{scorollary} {\bf [conformal class dense].}
 Let $M$ be a closed smooth $n$-manifold.
 Then each conformal class is dense in the moduli space of isometry classes of Riemannian metrics on $M$,
  endowed with the Gromov's $\varepsilon$-approximation topology.
\end{scorollary}

\bigskip

The whole Sec.\ 2 is devoted to the proof of Therem~2.1.

\bigskip
	
\subsection{Proof of the theorem}

We shall now construct a positive smooth function $\Omega$ on $M$
 such that there exists an $\varepsilon_0$-approximation
 $$
    R_{\varepsilon_0}\; :\; (M,\Omega^2ds_1^2)\; \longrightarrow\; (M, ds_0^2)\,.
 $$
The idea behind the construction is illustrated in {\sc Figure}~2-1-1.
The various $\delta$'s that appear in the proof are small positive numbers to be adjusted in the end.
The proof of two lemmas, Lemma 2.1.1 and Lemma 2.1.2, in the main line of the argument
 are left to Sec.~2.2 and Sec.~2.3 respectively to avoid distraction.
  %

 \begin{figure} [htbp]
  \bigskip
  \centering

  \includegraphics[width=0.8\textwidth]{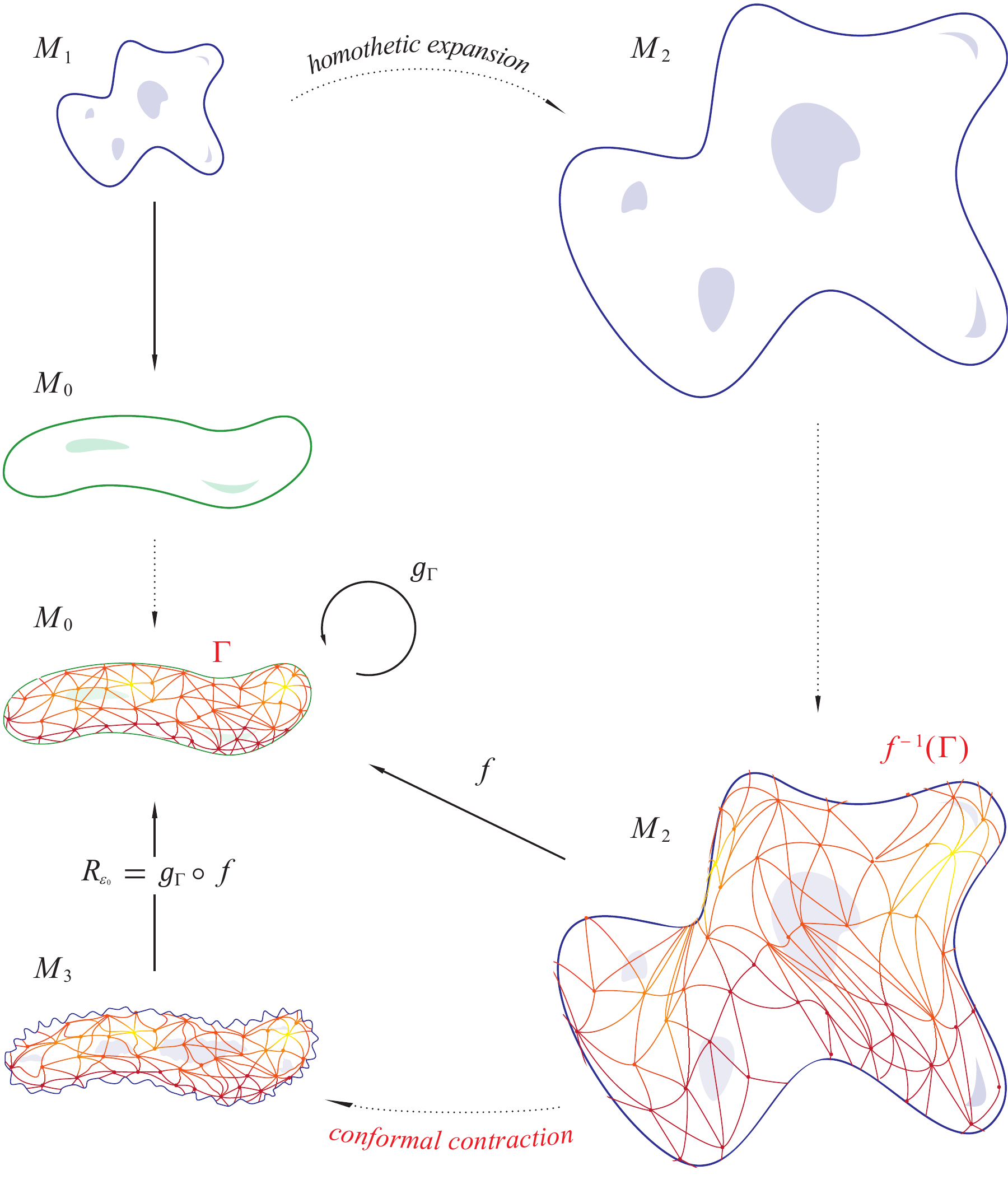}

  \bigskip
  \bigskip
  \centerline{\parbox{13cm}{\small\baselineskip 12pt
   {\sc Figure}~2-1-1.
     A sufficiently large homothetic expansion of $M_1$
	   followed by a conformal contraction guided by the pull-back graph $f^{-1}(\Gamma)$
	   of a dense enough geodesic graph $\Gamma\subset M_0$
	   leads to a new $M_3$ that is conformal to $M_1$ while being $\varepsilon_0$-close to $M_0$
	   in Gromov's $\varepsilon$-approximation topology.	
       }}
  \bigskip
 \end{figure}

\bigskip

Let $M_i$ be the Riemannian manifold $(M, ds_i^2)$, $i=0,\,1$.
The metric tensor $ds_i^2$ induces
   a norm  $\|\cdot\|_i$ on the fibers of the tangent bundle $T_{\ast}M$  of $M$ and
   a distance function $d_i$ on $M$.
Recall that for a Lipschitz map $f:M_i\rightarrow M_j$, its dilatation $\dil f$ is defined to be
 $$
   (\dil f)(p)\;:=\; \max_{X\in T_pM}\frac{\|f_{\ast}X\|_j}{\|X\|_i}
 $$
 for $p\in M$.
Now fix a diffeomorphism
 $$
   f\; :\; M_1\; \longrightarrow\;  M_0\,.
 $$
Given $0<\delta_1\ll 1$, let $C>0$ be large enough
 such that $f$ as a map
 $$
   f\;:\;  M_2:=(M, C^2ds_1^2)\; \longrightarrow\; M_0
 $$
 satisfies
 $$
    \dil f\; <\; \delta_1\; \ll\; 1\,.
 $$
Geometrically this means that by appropriately expanding $M_1$ homothetically to some $M_2$,
 one can make $f$ a highly contracting map.
 
\bigskip
 
\begin{lemma} {\bf [approximation by path-metrized graph].}
 Given any $\delta_2>0$, there exists a finite geodesic graph $\Gamma$ embedded in $M_0$
 such that its preimage $f^{-1}(\Gamma)$ in $M_2$ is $\delta_2$-dense and that
 for any pair of points $p$, $q$ in $\Gamma$,
  $$
     |d_{\Gamma}(p,q)- d_0(p,q)|\; <\; \delta_2\,,
  $$
  where $d_{\Gamma}$ is the edge-path distance function on $\Gamma$ induced from $ds_0^2$.
\end{lemma}
 
\medskip

\begin{lemma} {\bf [lining up least-contracted direction with graph].}
 For $\delta_1$ small enough
      in a way that depends only on $f$ and the homothety class of $M_0$ and $M_1$,
 given any $\delta_2$ and $\Gamma$ as in Lemma~2.1.1,
 there exists a Lipschitz homeomorphism $g_{\Gamma}:M_0\rightarrow M_0$ such that
  \begin{itemize}
   \item[(1)]
    $g_{\Gamma}$ fixes $\Gamma$ and is smooth around a neighborhood of $\Gamma$;
	
   \item[(2)]
    the composition $h:= g_{\Gamma}\circ f: M_2\rightarrow M_0$
	 remains a contracting map;
  
   \item[(3)]
     $\dil h$ at points in $f^{-1}(\Gamma)$ can be evaluated along the edge directions
	   except in a neighborhood of the set of vertices
   	    that consists of prongs of total length bounded above by some prescribed $\delta_3$.
    In other words, up to a small neighborhood of vertices, $h$ contracts the least along the edge directions
      for points in $f^{-1}(\Gamma)$.	
  \end{itemize}
\end{lemma}
  
\bigskip

\noindent
We will see how small $\delta_1$ should be from the proof of this lemma (cf.\ Sec.\ 2.3.4).

Let $A$ be the closure of the complement in $f^{-1}(\Gamma)$ of the collection of prongs in Lemma 2.1.2.
It is a collection of disjoint smooth arcs in $M$.
Let $U$ be an open set in $M_2$ that covers $A$ and over which the restriction of $h$ is smooth.
By the same argument as that in the existence of a partition of unity for a paracompact smooth manifold,
 there exists a smooth positive function $\chi$ on $M_2$ supported in $U$
  with values in the interval $[0,1]\subset {\Bbb R}$
  such that its restriction to $A$ is constant $1$.
Now let
 $$
   \Omega^{\prime}\; :=\; \chi\cdot \dil h\,+\, (1-\chi)\cdot 1\,;
 $$
then $\Omega^{\prime}$ is a positive smooth function on $M_2$ that satisfies
 $$
    \dil h\; \le \; \Omega^{\prime}\le 1\,.
 $$
 
Let
 $$
    \Omega\; :=\; \Omega^{\prime}C
	  \hspace{2em}\mbox{and}\hspace{2em}
	M_3\;:=\; (M, \Omega^2ds_1^2)\,.
 $$
Then, due to the above inequality for $\Omega^{\prime}$,
 $h$ is a non-expanding Lipschitz homeomorphism from $M_3$ to $M_0$.
Furthermore,
 let $d_{f^{-1}(\Gamma)}$ be the edge-path distance function on $f^{-1}(\Gamma)$
 induced from $\Omega^2ds_1^2$;
then due to Property (3) required of $h$ in Lemma 2.1.2,
 $h$ is length-preserving when restricted to the set of arcs in $A$ and the homeomorphism
  $$
     h|_{f^{-1}(\Gamma)}\;:\;
	    \left( f^{-1}(\Gamma)\,,\, d_{f^{-1}(\Gamma)}  \right)\;
		\longrightarrow\; (\Gamma, d_{\Gamma})
  $$
 is a $2\delta_3$-approximation.
 
Let $d_3$ be the distance function on $M$ induced $\Omega^2ds_1^2$.
One then has
 $$
    d_0(h(p), h(q))\; \le \; d_3(p,q)\; \le\;
	d_0(h(p), h(q))\,+\, (5\delta_2+2\delta_3)
 $$
 for any pair of points $p,\,q\in M_3$.
The first inequality follows from the fact that $h$ is non-expanding,
while the second inequality follows from the estimates :
 $$
  \begin{array}{ccl}
   d_3(p,q) & \le & d_{f^{-1}(\Gamma)}(p^{\prime}, q^{\prime})\,+\, 2\delta_2\,, \\[1.2ex]
      && \hspace{2em}
	        \mbox{where $p^{\prime}$ and $q^{\prime}$ are points in $f^{-1}(\Gamma)$
	          that are closest, relative to $d_3$, to $p$}\\
      && \hspace{2em}			
     	    \mbox{and $q$ respectively;} \\[1.2ex]
      & \le & [d_{\Gamma}(h(p^{\prime}), h(q^{\prime}))\,+\, 2\delta_3]\, +\, 2\delta_2\,, \\[1.2ex]
	  && \hspace{2em}
	        \mbox{since $h$ is a $2\delta_3$-approximation between the metrized graphs in question;}\\[1.2ex]
	  & \le & [d_0(h(p^{\prime}), h(q^{\prime}) )\,+\, \delta_2]\, +\, (2\delta_2+2\delta_3)\,,
            	    \hspace{2em}\mbox{by Lemma~2.1.1;}\\[1.2ex]
	  & \le & [d_0(h(p), h(q))\,+\, 2\delta_2]\, +\, (3\delta_2\,+\, 2\delta_3)\\[1.2ex]
	  && \hspace{2em}
	        \mbox{since both $d_0(h(p), h(p^{\prime}))$ and $d_0(h(q), h(q^{\prime}))$
	                         are less than $\delta_2$;}\\[1.2ex]
	  & =   & d_0(h(p), h(q))\,+\, (5\delta_2\,+\, 2\delta_3)\,.
  \end{array}
 $$
This shows that $h$ is a $(5\delta_2+2\delta_3)$-approximation from $M_3$ to $M_0$.
Since
  $M_3$ is conformally equivalent to $M_1$  and
  both $\delta_2$ and $\delta_3$ can be made arbitrarily small as long as $\delta_1$ is small enough,
 we conclude the theorem.
 
\noindent\hspace{40.8em}$\square$

\bigskip

\begin{remark} $[$picture behind the proof$\,]$. \rm
 The resulting $M_3$, though conformal to $M_1$, is in general extremely bumpy
   in the directions transverse to $f^{-1}(\Gamma)$.
 This very property makes $f^{-1}(\Gamma)$ a most efficient expressway in $M_3$ and, hence,
  captures the metrical characters of $M_3$ at scales larger than the size of the bumps.
 On the other hand, $f^{-1}(\Gamma)$ with the edge-path distance from $M_3$ is close to $\Gamma$,
   which in turn is close to $M_0$, in the sense of Gromov's topology.
 This is the geometric picture behind the proof. (Cf.\ {\sc Figure}~2-1-1.)
\end{remark}

\bigskip

\subsection{Proof of Lemma 2.1.1}

Since $M$	is compact, there exists a $c>0$ such that
 $$
    \frac{\|f_{\ast}X\|_0}{\|X\|_2}\; \ge c
 $$
 for all $X\in T_{\ast}M$.
Let $V_0$ be a $c\delta_2$-dense finite subset in $M_0$.
Attach to each pair of points in $V_0$ a minimal geodesic arc in $M_0$;
 we then obtain a finite geodesic graph $\Gamma_0\subset M_0$
  such that $f^{-1}(\Gamma_0)$ is $\delta_2$-dense in $M_2$.
For each edge $E$ of $\Gamma_0$, let
 $$
    U_E\; :=\; \cup_{p\in E}B(p;c\delta_2)
 $$
 be the $c\delta_2$-neighborhood of $E$ in $M_0$,
 where $B(p;c\delta_2)$ is the ball at $p$ of radius $c\delta_2$ in $M_0$.
For each $v_0\in V_0$ that lies in $U_E$,
 connect $v_0$ to $E$ by a minimal geodesic arc that realizes the distance $d_0(v_0, E)$.
Repeating this for all the edges $E\subset \Gamma_0$,
 we then obtain an enlarged geodesic graph $\Gamma$ from $\Gamma_0$.
We claim that $\Gamma$ satisfies the required properties.
 
First, it is clear that $f^{-1}(\Gamma)$ is $\delta_2$-dense in $M_2$
 since its subgraph $f^{-1}(\Gamma)$ already is.
Next, observe that $V_0$ is $3c\delta_2$-dense in $\Gamma$
 with respect to the edge-path distance $d_{\Gamma}$ on $\Gamma$.
Now let
  $p$, $q$ be any pair of points in $\Gamma$  and
  $v_1$ , $v_2$ be points in $V_0$ that are closest, with respect to $d_{\Gamma}$,
      to $p$ and $q$ respectively.
Then
 $$
   d_0(p,v_1)\; \le\; d_{\Gamma}(p,v_1)\; \le\; 3c\delta_2
    \hspace{2em}\mbox{and}\hspace{2em}
   d_0(q,v_2)\; \le\; d_{\Gamma}(q,v_2)\; \le\; 3c\delta_2\,.	
 $$
Thus
  $$
    \begin{array}{ccl}
	 d_0(p,q)& \le  &  d_{\Gamma}(p,q) \\[1.2ex]
	   & \le  & d_{\Gamma}(v_1,v_2)\,+\, 2(3c\delta_2)  \;
                       =\;    d_0(v_1,v_2)\,+\, 6c\delta_2   \\[1.2ex]
	   & \le  & [d_0(p,q)+2(3c\delta_2)]\, +\, 6c\delta_2\;
        	            =\; d_0(p,q)\,+\, 12c\delta_2\,.
    \end{array}
  $$
Consequently,
  $$
     |d_{\Gamma}(p,q)- d_0(p,q)|\;\le 12c\delta_2 \;\le \; \delta_2
  $$
 as required if one chooses $c$ small enough.
This concludes the lemma.
 
\noindent\hspace{40.8em}$\square$

\bigskip
  
Pictorially, $\Gamma$ is a net of straight expressways in $M_0$ with densely distributed entrance-'n'-exits,
 through which one can shift from one expressway to another efficiently for a shorter distance in the travel
 on $M_0$.

\bigskip

\subsection{Proof of Lemma 2.1.2}
	
The proof is divided into a sequence of steps which we now proceed.    	

\bigskip

\subsubsection{Basic bounds and Euclidean approximations}

Recall the fixed diffeomorphism
  $$
      f\;:\; M_1\,:=\, (M, ds_1^2)   \; \longrightarrow\; M_0\, :=\, (M, ds_0^2) \,.
  $$
Since $M$ is compact,
 there exists a $\kappa_0>0$ and an angle $\theta_0\in (0,\frac{\pi}{2})$
 such that
   $$
     \frac{\min_{X\in T_{\ast}M}\frac{\|f_{\ast}X\|_0}{\|X\|_1}}
	  {\max_{X\in T_{\ast M}}\frac{\|f_{\ast}X\|_0}{\|X\|_1}}\;
	   >\; \kappa_0
   $$
 and that for all pairs of $X$, $Y\in T_{\ast}M$ orthogonal to each other,
  the absolute angle $\angle(f_{\ast}X, f_{\ast}Y)$ of $f_{\ast}X$ and $f_{\ast}Y$
   is bounded by
   $$
     \theta_0\;<\;    \angle(f_{\ast}X, f_{\ast}Y)\; <\; \pi-\theta_0\,.
   $$
 Notice that the same bounds $\kappa_0$ and $\theta_0$  work for $f$
   independent of any homothetic change of the metric on $M_0$ or $M_1$.
Thus, both inequalities hold for $f$ from $M_2:=(M, C^2ds_1^2)$ to $M_0$ as well,
 no matter what  $C>0$  is chosen.
As we shall see,
 the dilatation of functions that appear in the following construction
   can be completely controlled by $\kappa_0$ and $\theta_0$.
  
Since $\Gamma$ is made up of disjoint geodesic arcs after a truncation of a small neighborhood of vertices,
 we shall discuss the local construction around a geodesic arc first.
Let
 $$
   \gamma\;:\; [0,l]\; \longrightarrow\; M_0
 $$
  be a simple geodesic arc parameterized by the arc-length.
Given a small $\eta_1>0$,
 let  ${\cal D}$ be a transverse smooth codimension-$1$ distribution along $\gamma$
  whose  unit normal vector field $\hat{n}$ satisfies that
  $$
    \angle(\hat{n}(t),\dot{\gamma}(t))\; \in\; [0,\theta_0)
	  \hspace{2em}\mbox{for all $t\in[0,l]$}
  $$
 and that
  $$
    \mbox{$\hat{n}(t)\;=\; \dot{\gamma}(t)\;$
	  in the $\eta_1$-neighborhood of the end-points $t=0$, $t-l$}\,,
  $$
 where $\dot{\gamma}$ is the tangent vector field along $\gamma$.
Recall $n$, the dimension of $M$.
Let
  $\bar{B}(0;R)$ be the closed ball of radius $R$
    at the origin of the $(n-1)$-dimensional Euclidean space with polar coordinates $(r,\Theta)$
and let
 $$
   \Delta_{(l,R)}\; :=\;  [0,l] \times \bar{B}(0,R)
 $$
  be the cylinder specified by $(l,R)$ with the product Euclidean metric.
Fix an orthonormal frame $\{e_i\}_{i=1}^n$ along $\gamma$ with $e_1=\dot{\gamma}$.
For $R$ small enough,
 recall then that
  one has a cylindrical coordinate system for a tubular neighborhood $K$ of $\gamma$ given by
  $$
    \begin{array}{cccccccc}
    \Exp_{\gamma} & :   & \Delta_{(l,R)}  & \stackrel{\sim}{\longrightarrow}
	     & K     & \hookrightarrow & M_0\\[1.2ex]
     && (t,r,\Theta)		 	 & \longmapsto              & \exp_{\gamma(t)}(r,\Theta)  &&&,
	\end{array}
  $$
 where one identifies $(t,r,\Theta)\in \Delta_{(l,R)}$
   to a point in $T_{\gamma(t)}M_0$ via $\{e_i\}_{i=1}^n$
   before taking the exponential map.
Notice that $\Exp_{\gamma\ast}$ is an isometry
  from $\left.T_{\ast}\Delta_{(i,R)}\right|_{\{r=0\}}$
  to $T_{\ast}M_0|_{\gamma}$.
Thus, given any $\eta_2>0$, as long as $R$ is small enough, one has
 $$
   \left|
      \frac{\|\Exp_{\gamma\ast}X\|_0}{\|X\|_E}\, -\, 1
   \right|  \; <\; \eta_2
 $$
 for all $X\in T_{\ast}\Delta_{(l,R)}$,
   where $\|X\|_E$ is the Euclidean norm of $X$.

\bigskip

\subsubsection{A Lipschitz anti-tilt homeomorphism on $\Delta_{(l,R)}$}

Let ${\cal H}$ be the codimension-$1$ distribution along $\{r=0\}$
  associated to the coordinate-hyperplane field $\{t= \mbox{constant}\}$.
We shall show that there exists a Lipschitz {\it anti-tilt} homeomorphism
 $$
   \tau\;:\; \Delta_{(l,R)}\; \longrightarrow\;  \Delta_{(l,R)}
 $$
 that satisfies the following conditions :
  \begin{itemize}
    \item[(1)]
	 The dilatation of $\tau$ is bounded above by a constant $C_{\theta_0}$
   	  that depends only on $\theta_0$.
	
	\item[(2)]
     $\tau$ fixes the core $\{r=0\}$	 and is smooth in a neighborhood of the core.
	
	\item[(3)]
     $\tau$ is the identity map in a neighborhood of the boundary $\partial\Delta_{(l,R)}$	
	   of $\Delta_{(l,R)}$.
	
	\item[(4)]
     $\tau_{\ast}|_{r=0}$	sends the pushed-forward distribution
	   $\left(\Exp_{\gamma}^{-1}\right)_{\ast}{\cal D}$ along $\{r=0\}$ to ${\cal H}$.
  \end{itemize}
An explicit example of such $\tau$ is given below.
  
Let $0<r_1<r_2<R$.
For $r_1$ small enough, we may identify the distribution
 $\left(\Exp_{\gamma}^{-1}\right)_{\ast}{\cal D}$
 with a smooth codimension-$1$ foliation in $\Delta_{(l, r_1)}$
  via the exponential map from the Euclidean metric.
This foliation is transverse to the coordinate lines that are parallel to the core of $\Delta_{(l,r_1)}$.
One can extend this foliation to a Lipschitz foliation ${\cal F}$ in $\Delta_{(l,R)}$
 by defining its leaf at $(t,r)=(t_0,0)$ to be the Lipschitz submanifold ${\cal F}_{t_0}$ :
 $$
   {\cal F}_{t_0}\; =\; \left\{
        \begin{array}{ccl}
		   \left(\Exp_{\gamma}^{-1}\right)_{\ast}{\cal D}_{t_0}
		       && \mbox{for region $\,\{r\le r_1\}$ }\\[1.2ex]
		 {\cal H}_{t_0}
		       && \mbox{for region $\,\{r_2\le r\le R\}$}\\[1.2ex]
		      \begin{array}{l}
			     \hspace{-1.5ex}\mbox{\LARGE $\cdot$}\:
		           \mbox{radial-linear interpretation between}\\[-.2ex]
				   \mbox{the above inner- and outer-pieces}
		      \end{array}
			   && \mbox{for region $\,\{r_1\le r \le r_2\}$}
        \end{array}
		                                                  \right.,
 $$
(cf.\ upper left in {\sc Figure}~2-3-2-1),
 where we identify ${\cal H}_{t_0}$ with the hyperplane $\{t=t_0\}$ in $\Delta_{(l,R)}$.
Regard $\Delta_{(l,R)}$ as an $I$-bundle (i.e.\ interval-bundle) over $\bar{B}(0;R)$.
Let
 $$
   \tau\, =\, \tau_{(r_1,r_2)}\;  =\;
    (\Delta_{(l,R)}, {\cal F})\; \longrightarrow\;  (\Delta_{(l,R)},{\cal H})
 $$
 be the unique $I$-bundle map
    that covers the identity map on $\bar{B}(0;R)$,
	       fixes the core of $\Delta_{(l,R)}$,   and
		   takes the leaves of ${\cal F}$ to leaves of ${\cal H}$,
cf.\ {\sc Figure}~2-3-2-1.
     %
%
 \begin{figure} [htbp]
  \bigskip
  \centering

  \includegraphics[width=0.8\textwidth]{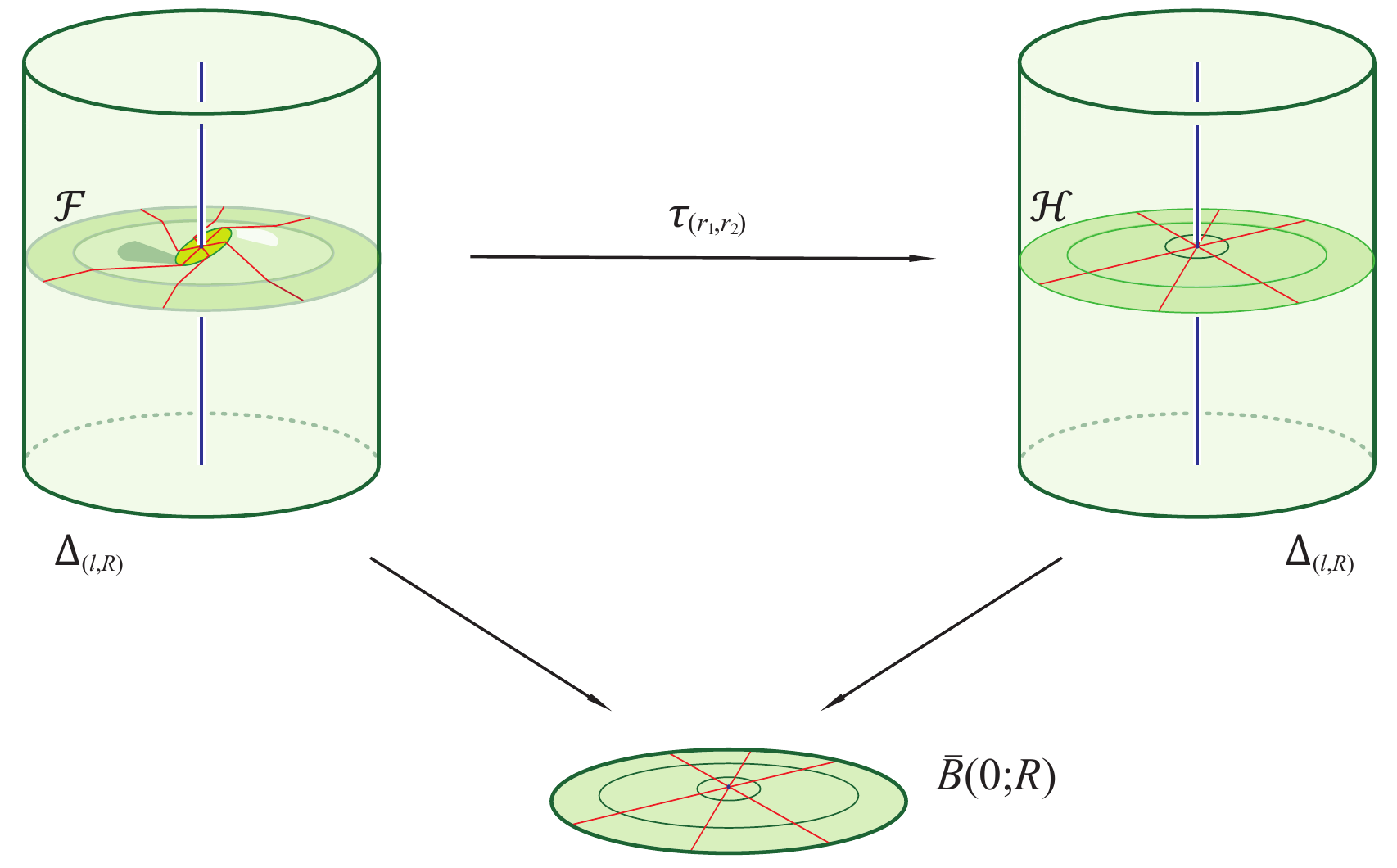}

  \bigskip
  \bigskip
  \centerline{\parbox{13cm}{\small\baselineskip 12pt
   {\sc Figure}~2-3-2-1.
        A Lipschitz {\it anti-tilt} homeomorphism $\tau$ on $\Delta_{(l,R)}$.
		Here we  treat $\Delta_{(l,R)}$ as an $I$-bundle over an $(n-1)$-disk
		   and $\tau$ as a bundle map.
		The behavior of $\tau$ on a generic disk-section is indicated.
		Three pairs of radial rays (in red) on these disks are added to help to illuminate further
		  how $\tau$ ``corrects" the tilted $(\Exp_{\gamma}^{-1})_{\ast}{\cal D}$
		  to ${\cal H}$ along the core (in blue) of $\Delta_{(l,R)}$.
       }}
  \bigskip
 \end{figure}	
Then, by construction,
 Conditions (2), (3), and (4) are satisfied.
Notice that $\tau_{(r_1,r_2)}$ is smooth on each of the three regions and
 that $\tau_{(r_1,r_2)\ast}$ can be made arbitrarily close to $\Id_{\ast}$
    on the region $\{r_1\le r\le r_2\}$ as long as $r_1$ and $R-r_2$ are small enough.
Thus,	by considering the bound for $\dil_{\tau_{(r_1,r_2)}}$ at the core,
  which can be worked out by the Lagrange multiplier method,
 one has
  \begin{eqnarray*}
   c_{\theta_0} & :=
        & \frac{1}{2}\,\sqrt{1\,+\, \frac{\cot^2\theta_0\,-\,\cot\theta_0\,\sqrt{4\,+\,\cot^2\theta_0}}
		                                                                    {2}}\\[1.2ex]
     & < & \dil\tau_{(r_1,r_2)} \\
	 & < & C_{\theta_0}\;
	             :=\;  2\,\sqrt{1\,+\, \frac{\cot^2\theta_0\,+\,\cot\theta_0\,\sqrt{4\,+\,\cot^2\theta_0}}
		                                                      {2}}\;,	
  \end{eqnarray*}
 for $r_1$, $R-r_2$ small enough.
Thus, Condition (1) can also be made satisfied.

\bigskip

\subsubsection{A Lipschitz radial-squeeze homeomorphism in $\Delta_{(l,R)}$}

Next we show that
  there exists a Lipschitz {\it radial-squeeze} homeomorphism
   $$
     \sigma\;:\; \Delta_{(l,R)}\; \longrightarrow\; \Delta_{(l,R)}
   $$
 that satisfies the following conditions :
 \begin{itemize}
  \item[(1)]
    The dilatation of $\sigma$ is bounded above by a constant $C_{\kappa_0,\theta_0}$
	  that depends only on $\kappa_0$ and $\theta_0$.
	
  \item[(2)]
   $\sigma$ fixes the core $\{r=0\}$ and is smooth in a neighborhood of the core.

  \item[(3)]
   $\sigma$ is the identity map in a neighborhood of the boundary $\partial\Delta_{(l,R)}$.

  \item[(4)]
   Recall ${\cal H}$ from Sec.\ 2.3.2 and $\eta_1$ from Sec.\ 2.3.1.
   Then,
    $\sigma_{\ast}|_{r=0}$ leaves ${\cal H}$ invariant and
	its dilatation in the radial directions satisfies
    $$
	   \frac{\|\sigma_{\ast}X\|_E}{\|X\|_E}\;
	      \le\; \frac{1}{2}\,\frac{\kappa_0}{C_{\theta_0}}
	$$
    for all $X$ in ${\cal H}|_{[\eta_1,\, l-\eta_1]}$.
 \end{itemize}
An explicit example of such $\sigma$ is given below.
 
Let
 $$
   \zeta\;:\; [0,l]\; \longrightarrow\; [0,1]
 $$
 be a smooth bump function defined on $[0,l]$ that takes constant value $1$ at $[\eta_1,l-\eta_1]$
   and $0$ in a neighborhood of the end-points.
Let $\lambda>0$.
Then there exists a smooth strictly increasing function
 $$
   \varrho\;:\; [0,R]\; \longrightarrow\; [0,R]
 $$
 that satisfies
   \begin{itemize}
     \item[\LARGE $\cdot$]
	   $\varrho(0)\,=\,0\,$, $\;0\,\le\, r-\varrho(r)\,<\,\lambda\;$  and
	   $\;\varrho(r)\,=\, r\,$ in a neighborhood of $R$.
	
	 \item[\LARGE $\cdot$]
	   $0\,<\, \varrho^{\prime}(0)\,\le\,\frac{1}{2}\frac{\kappa_0}{C_{\theta_0}}\;$
	     and
	   $\;\varrho^{\prime}(0)\,\le\, \varrho^{\prime}(r)\, <\, 1+\lambda\,$.
   \end{itemize}
   
Regard $\Delta_{(l,R)}$ as a $(n-1)$-disk-bundle over $[0,l]$.
Let
  $$
    \sigma\,=\, \sigma_{(\zeta,\varrho)}\,:\,
       \Delta_{(l,R)}\;  \longrightarrow\; \Delta_{(l,R)}
  $$
 be the disk-bundle map defined by
  $$
    \sigma_{(\zeta,\varrho)}(t,r,\Theta)\; =\;
	   (t\,,\,   \zeta(t)\varrho(r)\,+\, (1-\zeta(t))r\,,\,\Theta)\,,
  $$
  cf.\ {\sc Figure}~2-3-3-1.
    %
	%
 \begin{figure} [htbp]
  \bigskip
  \centering

  \includegraphics[width=0.8\textwidth]{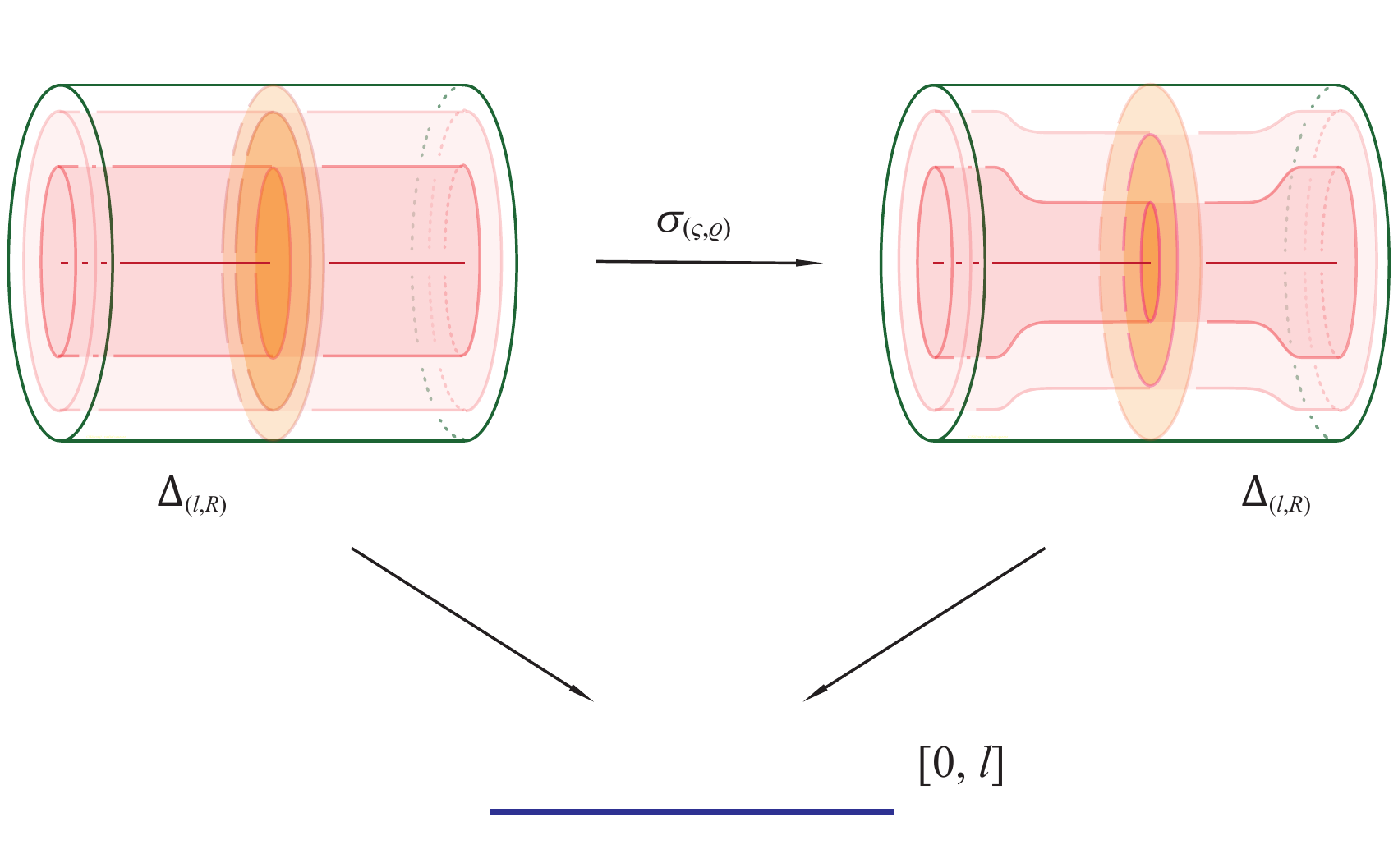}

  \bigskip
  \bigskip
  \centerline{\parbox{13cm}{\small\baselineskip 12pt
   {\sc Figure}~2-3-3-1.
        A Lipschitz {\it radial-squeeze} homeomorphism on $\Delta_{(l,R)}$.
       }}
  \bigskip
 \end{figure}	
Then, by construction,  Conditions (2), (3), and (4) are satisfied.
Straightforward computation shows that with respect to the cylindrical coordinate frame,
 the Jacobian matrix of $\sigma_{(\zeta,\varrho)}$ tends to a diagonal matrix
   with the diagonal entries bounded below by $\zeta^{\prime}(0)$ and above by $1$
  as $\lambda$ tends to $0$.
Consequently, an upper bound for $\dil \sigma_{(\zeta,\varrho)}$ can be lowered
 arbitrarily close to $1$ as long as $\lambda$ is small enough.
This shows that
  Condition (1) can be made satisfied
     with $C_{\kappa_0,\theta_0}$ any number greater than $1$
  if one chooses $\lambda$ sufficiently small.

\bigskip

\subsubsection{Conjugation back to $K$}

Let
 $$
    g_{(\gamma,K)}\;=\; \Exp_{\gamma}\circ(\sigma\circ \tau)\circ\Exp_{\gamma}^{-1}\;
	 :\; K\;  \longrightarrow\; K\,.
 $$
Recall $\eta_2$ from Sec.\ 2.3.1.
Then the dilatation of $g_{(\gamma,K)}$ satisfies
 $$
    \dil g_{(\gamma,K)}\;
	 <\;   \frac{1+\eta_2}{1-\eta_2}\,\cdot\, C_{\theta_0}\cdot C_{\kappa_0,\theta_0}\;
	 <\;  2\,C_{\theta_0}C_{\kappa_0,\theta_0}
 $$
 provided that $\eta_2$ is sufficiently small.

Notice that the last upper bound depends only on $\kappa_0$ and $\theta_0$,
  which, as one may recall from Sec.\ 2.3.1,
     depends only on $f$ and the homothety class of $M_0$ and $M_1$.
Thus, the same construction for $g_{(\gamma,K)}$ and
   upper bounds for $\dil g_{(\gamma,K)}$	 work as well for the triple $(M_2, M_0, f)$.
We assume therefore that
 the $\delta_1$ that appears in the beginning of the proof of Main Theorem satisfies
  $$
    \delta_1\cdot 2\,C_{\theta_0}C_{\kappa_0,\theta_0}\; <\; 1\,.
  $$
Then, the composition
  $$
     g_{\gamma,K}\circ f
  $$
 from some domain in $M_2$ to $M_0$ will always remain contracting.
Observe also that
 \begin{itemize}
   \item[(1)]
    $g_{(\gamma,K)}$ fixes $\gamma$ and is smooth in a neighborhood of $\gamma$;

   \item[(2)]	
    $g_{(\gamma,K)}$ is the identity map in a neighborhood of the boundary $\partial K$ of $K$;
  
   \item[(3)]
    $g_{(\gamma,K)\ast}$ sends the distribution ${\cal D}$ along $\gamma$
 	   to the distribution $\Exp_{\gamma\ast}{\cal H}$   and
    it satisfies 	
      $$
	     \frac{\|g_{(\gamma,K)\ast}X\|_0}{\|X\|_0}\;
		  <\; 2\cdot \frac{1}{2}\frac{\kappa_0}{C_{\theta_0}}\cdot C_{\theta_0}\,
		        =\; \kappa_0		
	  $$
	  for all $X\in {\cal D}$.
 \end{itemize}

\bigskip

\subsubsection{Application to $\Gamma$}

Truncate out from $f^{-1}(\Gamma)$
  a neighborhood of the set of vertices that consists of a collection of prongs of total length in $M_2$
   less than some prescribed $\frac{1}{2}\delta_3$.
Let $A^{(e)}$ be the closure of the remaining part.
A component $A_i$ of $A^{(e)}$ is simply a minutely reduced edge of $f^{-1}(\Gamma)$  and
 the image $f(A^{(e)})$ is a finite collection of disjoint simple geodesic arcs
   $\gamma_i:=f(A_i)$ in $M_0$, parameterized by the arc-length.
Let ${\cal E}_i$ be the orthogonal complement of the tangent line field to $A_i$ in $M_2$.
Then $f_{\ast}{\cal E}_i$ is a codimension-$1$ transverse distribution along $\gamma_i$
 that satisfies the angle condition
  $$
     \angle(\hat{n}_i,\dot{\gamma}_i)\; \in \; [0,\theta_0)
  $$
 in Sec.\ 2.3.1,
    where $\hat{n}_i$ is the unit normal vector field to $f_{\ast}{\cal E}_i$
	that has positive inner product with $\dot{\gamma}_i$.
Modify $f_{\ast}{\cal E}_i$ smoothly in a small neighborhood of the end-points of $\gamma_i$	
 so that for the resulting distribution ${\cal D}_i$,
   the angle condition still holds   and
   $\hat{n}_i=\dot{\gamma}_i$ around the end-points.
We assume that
 the total length of these small neighborhood of end-points of $A_i$'s is less than $\frac{1}{2}\delta_3$.

Choose a collection of disjoint tubular neighborhood $K_i$ around each $\gamma_i$
 so that  the inequality for the dilatation of $\Exp_{\gamma_i}$ in Sec.\ 2.3.1 holds for all $i$.
Applying the argument in Sec.\ 2.3.2 --- Sec.\ 2.3.4,
 one then has, for each $(\gamma_i,K_i)$, a Lipschitz  homeomorphism
  $$
     g_{(\gamma_i,K_i)}\; :\; K_i\; \longrightarrow\; K_i\,,
  $$
 as constructed previously for general geodesic arcs.	
Its composition with $f$, as a map from a domain in $M_2$ to $M_0$, has dilatation
 $$
   \dil(g_{(\gamma_i, K_i)}\circ f)\;
     <\; \delta_1\cdot 2\,C_{\theta_0}C_{\kappa_0,\theta_0}\; <\; 1	
 $$
 and
 $$
   \frac{\|(g_{(\gamma_i,K_i)}\circ f)_{\ast}Y\|_0}{\|Y\|_2}\;
    <\;  \kappa_0 \cdot
	          \frac{\|f_{\ast}Y\|_0}{\|Y\|_2}\;
    <\;  \min_{Y^{\prime}\in T_{\ast}M}
	          \frac{\|f_{\ast}Y^{\prime}\|_0}{\|Y^{\prime}\|_2}\;
    <\;  \frac{\|f_{\ast}\dot{\gamma}_i(t)\|_0}{\|\dot{\gamma}_i(t)\|_2}
 $$
  for any $t$ and for any $Y\in {\cal E}_i$ except in a small neighborhood of the end-points of $A_i$.
Consequently, for points on the arc $A_i$,
    $(g_{(\gamma_i,K_i)}\circ f)_{\ast}$ contracts the least along the tangent directions to $A_i$
	except in a small neighborhood of its end-points.
	
Since the restriction of $g_{(\gamma_i,K_i)}$ to a neighborhood of the boundary $\partial K_i$
 of its defining domain $K_i$	is the identity map,
 one can extend
  $$
   \coprod_i g_{(\gamma_i,K_i)}\; :\;
     \coprod_i K_i\; \longrightarrow\; \coprod_i K_i
  $$
 to a Lipschitz homeomorphism
  $$
    g_{\Gamma}\;:\; M_0\; \longrightarrow\; M_0
  $$
  on the whole $M_0$ by setting $g_{\Gamma}$ to be the identity map
    outside $\coprod_iK_i$.
The properties of $g_{(\gamma_i,K_i)}$ imply that
  the requirements for $g_{\Gamma}$ and $h:=g_{\Gamma}\circ f$ in Lemma~2.1.2
  are satisfied.
   
\bigskip

This completes the proof of Lemma~2.1.2.

\bigskip

\noindent\hspace{40.8em}$\square$

\bigskip

\section{Another question -- Toward a ``Gromov-Wilson Theory"?}

Gromov's {\it geometry-at-large-scale} base on his topology reminds one
 of Wilson' s{\it renormalization group flow}.
We shall now give a sketch of the latter for the sake of completeness and pose a question on uniting both.

\bigskip

\begin{flushleft}
{\bf Wilson's renormalization group theory}
\end{flushleft}
In exploring the question
 \begin{itemize}
  \item[Q.] {\bf [Wilson]}$\,$ {\it
    How can one understand a physical system
	when its correlation length is large relative to the fundamental length?}
 \end{itemize}
 Kenneth Wilson developed the theory of {\it renormalization group}.
Abstractly and slightly mathematically rephrased,
 it consists of the following ingredients and procedures
 ([Al2], [Ca], [Go], [I-D], [LeB], [W-K]) :
\begin{itemize}
 \item[(1)]
  The {\it space ${\cal S}\times{\Bbb R}_+$ of all physical systems} of a certain class
  {\it with a fundamental length},
   governed by either Lagrangian or Hamiltonian densities of a certain type.
  One parameterizes ${\cal S}\times{\Bbb R}_+$ by
   $(g^1,\,g^2,\, \cdots\,;\,\varepsilon)$,
    where
	   $g^i$ are the coupling constants that appear in these densities and
	   $\varepsilon$ is the fundamental length corresponding to a cutoff that sets the length scale.
  Associated to this, one has the {\it dimension-depriving map}
    $$
	 \begin{array}{ccccc}
	  D_{pr} & : & {\cal S}\times{\Bbb R}_+ & \longrightarrow & {\Bbb R}^{\infty}\\[.6ex]
	  && (g^1,\,g^2,\,\cdots\,;\,\varepsilon)
	     & \longmapsto & (g^1\varepsilon^{-d_1},\, g^2\varepsilon^{-d_2},\,\cdots\,)\,,	
	 \end{array}
    $$	
    where $d_i$ is chosen so that $g^i\varepsilon^{-d_i}$ becomes dimensionless.
  We shall denote the image $\Image D_{pr}$ by ${\cal S}^{\dimless}$.
  
 \item[(2)]
  A {\it mode expansion} for fields $\phi$ in the theory with respect to length scales.
 Formally,
  $$
    \phi\;=\; \int_0^{\infty}\phi^{(\varepsilon)}\,d\varepsilon\,.
  $$
 Define the {\it truncated fields} by
  $$
   \phi_{\varepsilon}\;
     :=\; \int_{\varepsilon}^{\infty}\phi^{(\varepsilon^{\prime})}\,d\varepsilon^{\prime}\,.
  $$
 Given
   \begin{itemize}
     \item[\LARGE $\cdot$]
      $(g^1,\,g^2,\,\cdots\,;\,\varepsilon)\in {\cal S}\times{\Bbb R}_+$,
        representing a physical system at scale $\varepsilon$,  and
		
     \item[\LARGE $\cdot$]		
      $\lambda\ge 1$,
   \end{itemize}
   by integrating out the modes $\phi^{(\varepsilon^{\prime})}$ of fields $\phi_{\varepsilon}$
         with $\varepsilon^{\prime}\in [\varepsilon, \lambda\varepsilon)$
    in the path-integral
	$$
	   Z[g^1,\,g^2,\,\cdots\,;\,\varepsilon]\; :=\;
	      \int [{\cal D}\phi_{\varepsilon}]
		    e^{\frac{\sqrt{-1}}{\hbar}I_{(g^1,\,g^2,\,\cdots)}(\phi_{\varepsilon})}\,,
	$$
   one obtains a $(g^{\prime 1},\,g^{\prime 2},\,\cdots\,)\in {\cal S}$  	
    such that
	 $$
	    Z[g^1,\,g^2,\,\cdots\,;\,\varepsilon]\;
		 =\; Z[g^{\prime 1},\,g^{\prime 2},\,\cdots\,;\,\lambda\varepsilon] \,.
	 $$
   (A Wick-rotation is understood implicitly for the exponents if necessary.
       Also, when the theory is for fields on a lattice or a graph or when the fields take only finitely many values,
	   some modifications that replace $\int$ by appropriate $\sum$ are understood in these expressions.)
\end{itemize}
The map from ${\cal S}\times{\Bbb R}_+$ to itself given by
 $$
   (g^1,\,g^2,\,\cdots\,;\,\varepsilon)\;
      \longmapsto\; (g^{\prime 1},\,g^{\prime 2},\,\cdots\;;\;\lambda\varepsilon)	
 $$
 determines a {\it semi-flow} in ${\cal S}\times{\Bbb R}_+$
   (i.e.\ an $({\Bbb R}_{\ge 1},\times)$-action)
   whose trajectories are parameterized by $\lambda\in {\Bbb R}_{\ge1}$.
Theories in ${\cal S}\times{\Bbb R}_+$ that lie in the same trajectory of the flow are regarded
 as one theory being looked at at different length scales.
 Reparameterize these trajectories by $t=log\lambda$,
 they become descendable to
  ${\cal S}^{\dimless}$ via $D_{pr}$.
The result is the Wilson's {\it renormalization group (RG-)flow
   --- now an $({\Bbb R}_{\ge 0},+)$-action --- on ${\cal S}^{\dimless}$}.
The transformations $R_t$ it generates are the {\it RG-transformations}.
The components of the vector field on ${\cal S}^{\dimless}$ that generates the flow
  are called the {\it $\beta$-functions}.

Let $\xi_{\phys}(g^1,\,g^2,\,\cdots\,)$ be the physical correlation length of the system
 specified by $(g^1,\,g^2,\,\cdots)$.
Define the dimensionless correlation length function $\xi$ on ${\cal S}\times{\Bbb R}_+$ by
 $$
   \xi\;:\; (g^1,\,g^2,\,\cdots\,;\,\varepsilon )\; \longmapsto\;
     \xi_{\phys}(g^1,\,g^2,\,\cdots\,)/\varepsilon\,.
 $$
 Being dimensionless, this map descends to ${\cal S}^{\dimless}$ via $D_{pr}$.
We shall denote its descendant on ${\cal S}^{\dimless}$ still by $\xi$.
For any $P\in {\cal S}^{\dimless}$ and $t\ge 0$, one has that
 $$
   \xi(R_tP)\; =\; e^{-t}\xi(P)\,.
 $$
Following from this, if one lets $H_a$ be the level hypersurfaces $\xi^{-1}(a)$ of $\xi$,
 then the RG-flow is transverse to $H_a$'s and takes an $H_a$ to another $H_{a^{\prime}}$.
 \begin{itemize}
  \item[(3)] {\bf [Wilson's criterion]}$\,$
   Let $\xi_{\phys, 0}$ be the physical correlation length of the system specified by
    $(g_0^1,\,g_0^2,\,\cdots\,)$.
   Let
     $$
	   \gamma(t)\;=\; D_{pr}(g_0^1,\,g_0^2,\,\cdots\,;\, e^{-t}\xi_{\phys,0})
	     \hspace{2em}\mbox{and}\hspace{2em}
	   Q(t)\;=\; R_t \gamma(t)\,.	
     $$
   Then the curve $Q(t)$ lies in $H_1$.
   They corresponds to truncated theories
     whose correlation length and  fundamental length are equal  and
	     hence need much fewer degrees of freedom in describing them
	 than their ancestors backtracking along the RG-flows. 	 \\[.6ex]
	$\mbox{\hspace{1em}}$
    To make sure that
      \begin{itemize}
	   \item[($\dagger$)]
	     any curve $Q(t)$ in $H_1$ thus obtained has a limit as $t\rightarrow\infty$ and
	
	   \item[($\ddagger$)]
	    all such limits lie in a finite-dimensional submanifold ${\cal S}^{\dimless}_{\infty}$
		  of ${\cal S}^{\dimless}$,
      \end{itemize}	
	Wilson lays down some criteria concerning
	 {\it the fixed points and their stable and unstable manifolds of the RG-flow}
	  that together govern the dynamics of the flow.
   (E.g.\ the existence of hyperbolic fixed points with finite-dimensional unstable manifolds.)
   In the case of lattice field theories,
    when the flow has properties ($\dagger$) and ($\ddagger$),
	one can then take the truncated theory corresponding to the limit point
    	$Q(\infty)\in {\cal S}^{\dimless}_{\infty}$ as the {\it continuum theory}
		of the original discretized theory.
  With ${\cal S}^{\dimless}_{\infty}$ being finite-dimensional,
   the form of continuum theories can then be effectively predicted.
   (Cf.\ {\sc Figure} 3-1, adapted from [Al2] and [W-K].)
      %
 \end{itemize}

 \begin{figure} [htbp]
  \bigskip
  \centering

  \includegraphics[width=0.8\textwidth]{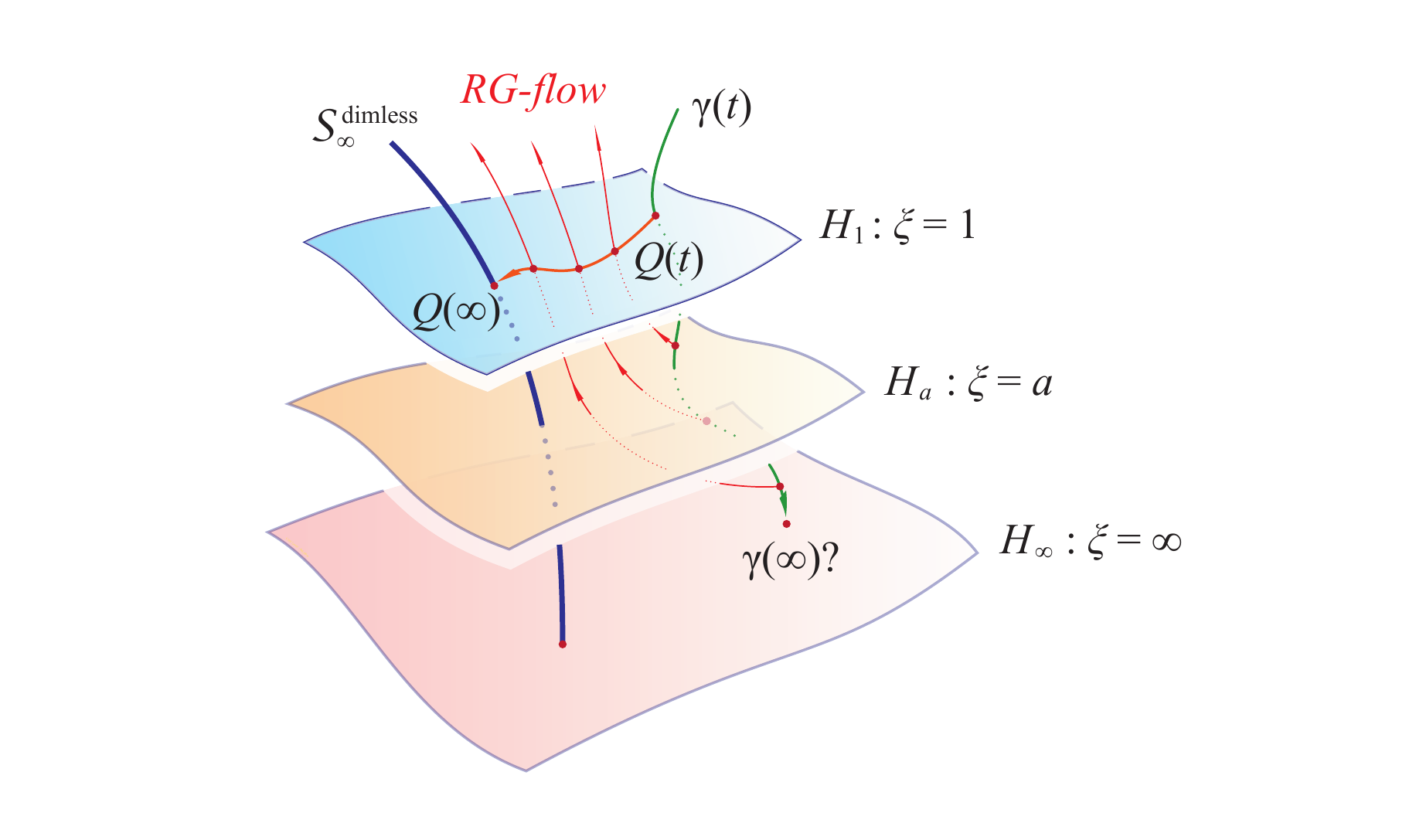}

  \bigskip
  \bigskip
  \centerline{\parbox{13cm}{\small\baselineskip 12pt
   {\sc Figure}~3-1.
        Through Wilson's RG flows, the limit system is taken in $H_1$, which then becomes manageable.
       }}
  \bigskip
 \end{figure}

\bigskip

\begin{flushleft}
{\bf Gromov's geometry-at-large-scale}
\end{flushleft}
Recall that associated to a set of generators $\{a_1,\,a_2,\,\cdots\,\}$ of a discrete group $G$,
 one can define a {\it word metric} $d_{\word}$
  as the unique maximal left invariant distance function thereon such that
   $d_{\word}(a_i,e)=d_{word}(a_i^{-1},e)=1$ for all $i$,
   where $e$ is the identity element of $G$.
To make the concepts from continuum geometry applicable to discrete groups,
 Gromov poses the following question :
 \begin{itemize}
  \item[\bf Q.] {\bf [Gromov]}$\,$ \it
   What does a metric space look like when being observed at a very large scale?
 \end{itemize}
He discovers that, when appropriately defined,
 the gemetry of the {\it large-scale limit} of $(G,d_{word})$ can reveal deeply properties of $G$ itself.
This is a key motivation for Gromov's topologies,
  under which a large-scale limit for a metric space $(X,d)$ is defined to be
  $\lim_{\lambda\rightarrow\infty}(X,\frac{1}{\lambda}d)$.
(Technically, one may need in addition an ultrafilter to obtain a limit.)
 
Turning to physics, taking continuum limit of discrete or lattice-like metric spaces
 are typical in statistical or lattice gauge field theories.

\bigskip

\begin{sexample} {\bf [taxi-cab metric as limit of metrized lattice].} \rm
 The large-scale limit of an infinite $4$-valence regular graph $\Gamma$ in the Euclidean plane
  with the induced edge-path distance function $d$ turns out to be the $2$-plane ${\Bbb R}^2$
  with the taxi-cab metric $d_{\infty}$ under $\varepsilon$-approximation topology
  (cf.\ {\sc Figure}~3-2).
    %
	%
 For a pair of points $p=(x_1,y_1), q=(x_2,y_2)\in {\Bbb R}^2$,
  $$
     d_{\infty}(p,q)\; =\; |x_1-x_2|+|y_1-y_2|\,.
  $$
 Note that as metric spaces,
  this is different from the Euclidean plane physicists tends to take as the continuum limit.
 
 \begin{figure} [htbp]
  \bigskip
  \centering

  \includegraphics[width=0.8\textwidth]{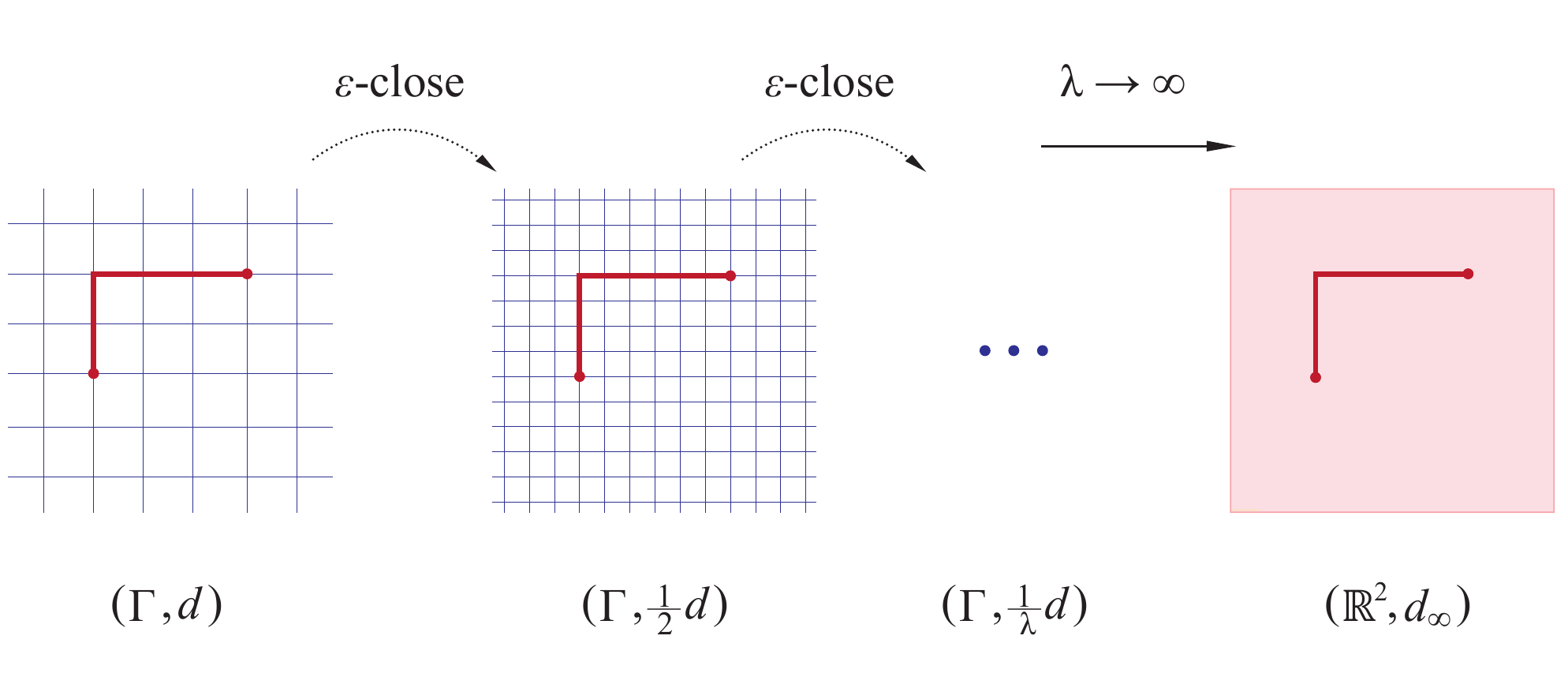}

  \bigskip
  \bigskip
  \centerline{\parbox{13cm}{\small\baselineskip 12pt
   {\sc Figure}~3-2.
        The limit of $(\Gamma, \frac{1}{\lambda}d)$, as $\lambda\rightarrow\infty$,
   		is the {\it taxicab geometry} on the $2$-plane.
       }}
  \bigskip
 \end{figure}	
\end{sexample}

\bigskip

Gromov's topologies make ``{\it taking the large-scale limit of metric spaces}" precise.
But the above example indicates that
  the limit may not be the same as what physicists expect.
In general, the large-scale limit of a metric space could be far from one's intuition.
(Try the standard hyperbolic $n$-space ${\Bbb H}^n$ or
     the Cayley graph of the fundamental group of a surface of genus $\ge 2$
	   with each edge assigned length $1$.)
Though the field-theorist's perception and the geometer's rigor may not coincide,
{\it isn't there a similar tune hidden in both Wilson's and Gromov's question?}

\bigskip

\begin{flushleft}
{\bf Wilson $\boldmath +$ Gromov ${\boldmath =\,?}$}
\end{flushleft}
In studying the effect under length rescaling,
 Wilson, as a physicist, takes the field-theoretical objects as primary concern,
 while Gromov. as a geometer, focuses on the underlying metric spaces.
It is natural to ask :
 \begin{itemize}
  \item[\bf Q.] {\bf [Gromov+Wilson]}$\,$ \it
     Can Gromov's and Wilson's view be merged into a unified one?
 \end{itemize}
Cf.\ {\sc Figure}~3-3.
           %
 
 \begin{figure} [htbp]
  \bigskip
  \centering

  \includegraphics[width=0.8\textwidth]{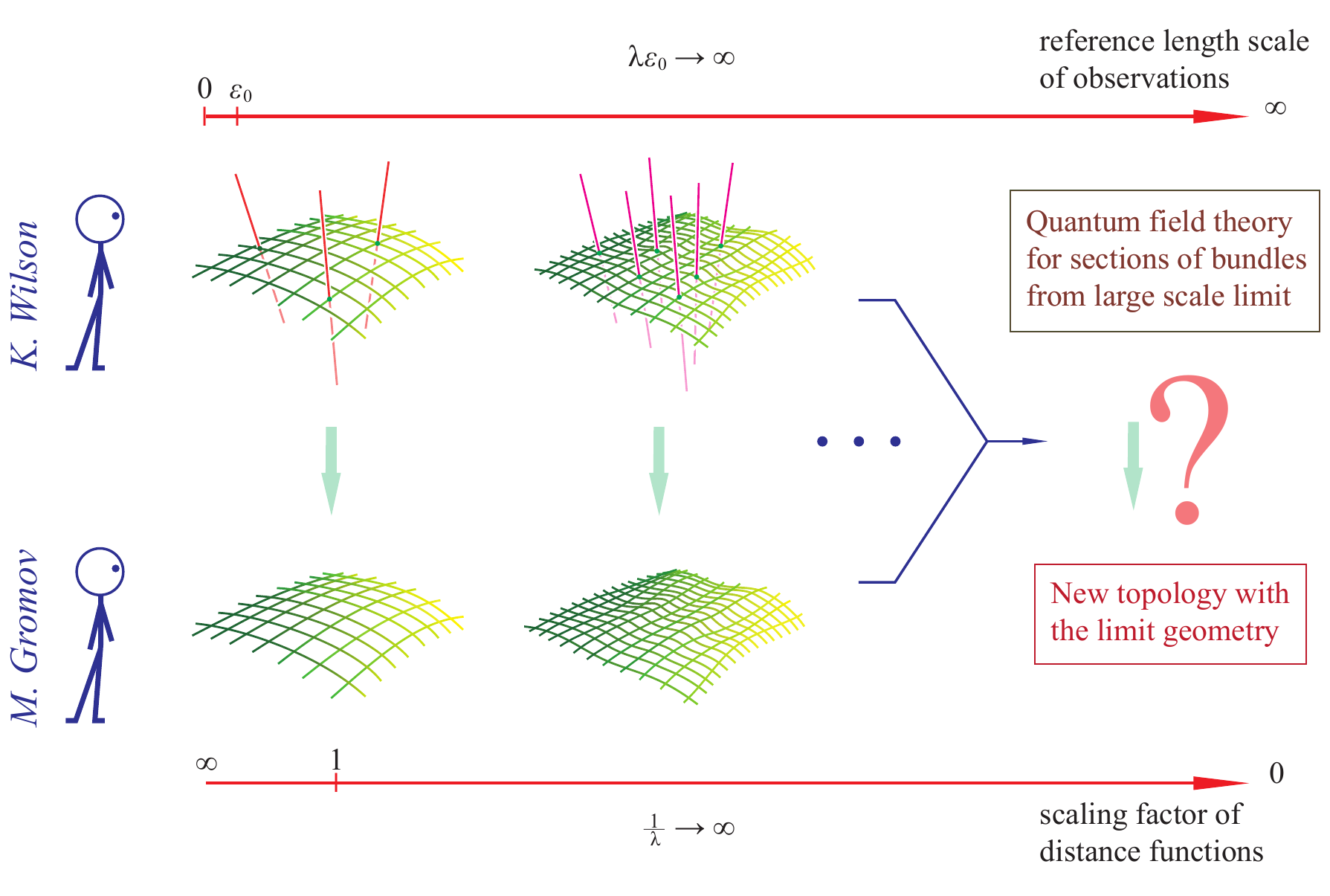}

  \bigskip
  \bigskip
  \centerline{\parbox{13cm}{\small\baselineskip 12pt
   {\sc Figure}~3-3.
        {\bf Q.} {\it Can Gromov's and Wilson's view merge into a unified one?}
		 A physical field is represented by a section of a bundle over a metric space.
		 Some fibers of the bundle are indicated.
       }}
  \bigskip
 \end{figure}

Among many sub-questions, let us address only a few.
\begin{itemize}
  \item[(1)]
   Taking the standard viewpoint from particle physics,
    fields are sections of various bundles $E$ over a metric space $(X,d)$.
  (In saying so, we let the gravitation field stand out separately as associated to the distance function $d$,
       which makes ``a reference length scale" acquire a well-defined meaning.)
    \begin{itemize}
	  \item[\bf Q1.]\it
	    Could one extend Gromov's topology to the space of isomorphism classes of bundles over  metric spaces
		so that one can define $\lim_{\lambda\rightarrow\infty}(E;X,\frac{1}{\lambda}d;s)$,
		where $s$ is a section of $E$?
    \end{itemize}	
    
  \item[(2)]
   Fixed points of Wilson's RG-flow corresponding to critical physical systems
    that are invariant under global rescaling.
   \begin{itemize}
    \item[\bf Q2.]\it
	 How should the physics of such critical system
	  and the large-scale invariants of its supporting metric space be related?
   \end{itemize}
   For example, could the anomalous dimension of a critical system be related to
   a certain large-scale dimension of its supporting metric space?
   Or could it indicate that in the prospect theory that accommodate both Gromov and Wilson
    a critical system may be supported by a self-similar fractal geometry?
   
  \item[(3)]
    Theorem 2.1 and Example 3.1 suggest that for a lattice to approximate a required continuum geometry,
	 one needs to consider also general conformal deformations.
	Such generalization of Wilson's theory has been investigated by physicists (e.g.\ [Po2], [Sc]).
	\begin{itemize}
	 \item[\bf Q3.]\it How could one take this into account on the way toward a synthesis of Gromov and Wilson? 	
    \end{itemize}
\end{itemize}

With these open questions, we conclude this note.

\vspace{12em}

\baselineskip 13pt
{\footnotesize

\vspace{1em}

\noindent
{\bf Notes added (2015).}
 The Main Theorem was proved sometime in 1990-1991 under Thurston's supervision.
 Since the previous revision in 1995,
  twenty years have passed and
  numerous string-theorists and geometers have crossed my path during this time,
  including, likely the most noticeable, Prof. Shing-Tung Yau,
   who and whose group work also on various mathematical issues on general relativity.
 In retrospect, while it is very attempting (and indeed very natural as well) to suspect a link between
  Gromov's work and quantum gravity and Wislon's renormalization group picture on a theory-space,
 in reality the physics language is more inclined to be (at least formally) analytical   and, thus,
 unless one is able to develop some version of analysis
    that fits well with Gromov's $\varepsilon$-approximation topology,
 it is extremely difficult to make any further progress on the problem addressed in this note.
This is the work mentioned in the dedication of [L-Y] to the memory of Prof.\ Thurston.
Except the overall re-typing, some footnotes, the correction of typos, the re-drawn figures,
     some re-editing, slight rephrasings and/or additional words to improve clarity, and two additional references,
  no changes are attempted/made to the original draft, 
  previously last revised in November 1995.

\bigskip

\noindent
\parbox[t]{5em}{[L-Y]}\
   \parbox[t]{44em}{C.-H.~Liu and S.-T.~Yau,
  {\it D-branes and Azumaya/matrix noncommutative differential geometry,
 I: D-branes as fundamental objects in string theory  and differentiable maps
    from Azumaya/matrix manifolds with a fundamental module to real manifolds},
  arXiv:1406.0929 [math.DG]. (D(11.1))}

\vspace{12em}

\noindent
{\bf Notes added after the first posting (October, 2015).}
(1)
The idea presented in this thesis:
  \begin{itemize}
    \item[\LARGE $\cdot$]\it
      equip Wheeler's superspace
       with Gromov's $\varepsilon$-approximation topology
      and use this ``master moduli space"\\
       as a tool/platform to study quantum gravity
  \end{itemize}
  came from a discussion with my then advisor Prof.\ William Thurston
  in fall, 1990, at his office on the third floor of Fine Hall
  at Princeton University.
 After my explanation to him Wheeler's interpretation of Einstein's vision
  of quantum gravity via geometrodynamics,
 Prof.\ Thurston looked struck with something and then started to explain
  to me on the blackboard Gromov's $\varepsilon$-approximation topology
  on the moduli space of isometry classes of metric spaces
   (i.e.\ topological spaces with a distance function)
  --- originally developed in part for the study of hyperbolic groups
   and their limits from the aspect of large-scale geometry.

 However, completely unaware to us and prior to Gromov's work,
 in the following work of David Edwards (1968),
 which is twenty-two years ahead of us:

\bigskip

\noindent
\parbox[t]{5em}{[Ed]}\
   \parbox[t]{44em}{D.A.\ Edwards,
  {\it The structure of superspace},
   in {\sl Studies in topology}, N.M.\ Stavrakas and K.R.\ Allen eds.,
   121--133,
   Academic Press, 1975.}

\bigskip

\noindent
a notion of {\it $\varepsilon$-isometry} was introduced
 ([Ed: Definition III.1]),
which has the same essence as Gromov's notion of $\varepsilon$-approximation.
It induces a topology on the moduli space $S$ of compact metric spaces.
Various basic properties of $S$ with this topology were studied ibidem
 ([Ed: Sec.\ III]).
This is likely the first work that intends to study Wheeler's superspace
 with such a topology that so naturally takes into account
 the fluctuations of both the topology and the geometry of a space
  (cf.\ [Ed: Sec.\ II]).
I thank the author for
 the communications (September, 2015) on his pioneering work
 after the previous posting of the thesis
  that updated my understanding of the history of the related development
 and sharing with me very generously his insights on the problem.

\bigskip

(2) The proof of the Main Theorem,
     which involves a homothetic expansion of $M_1$
      with a large constant conformal factor and then followed
      by a specifically designed conformal contraction of the resulting $M_2$ 
      (cf.\ {\sc Figure}~2-1-1),
    reminds one of an inflation scenario in cosmology.
    Whether there is truly a deep connection
      or an application of the Main Theorem to cosmology
     is unknown.

\vspace{3em}

\noindent
chienhao.liu@gmail.com, chienliu@math.harvard.edu

}

\end{document}